\documentstyle[12pt]{article}
\def\bq{\begin{equation}}
\def\eq{\end{equation}}
\def\bqa{\begin{eqnarray}}
\def\eqa{\end{eqnarray}}
\def\bqb{\begin{eqnarray*}}
\def\eqb{\end{eqnarray*}}
\def\eg{{\it e.g.\/}}

\global\nulldelimiterspace = 0pt

\def\Bsl{\hbox{/\kern-.6700em$B$}}    %  Bslash
\def\Dsl{\hbox{/\kern-.6700em$D$}}    %  Dslash
\def\Wsl{\hbox{/\kern-.6700em$W$}}    %  Wslash

\def\roughly#1{\mathrel{\raise.3ex
    \hbox{$#1$\kern-.75em\lower1ex\hbox{$\sim$}}}}
\def\lsim{\roughly<}
\def\gsim{\roughly>}

\def\ol#1{\overline{#1}}

\def\O{ {\cal O }}

\def\mz{M_Z}

\input tcilatex

\begin{document}

\pagenumbering{arabic} \thispagestyle{empty}

\begin{flushright} hep-ph/9509316\\
PM/95-30 \\ THES-TP 95/08 \\
September 1995 \\
Corrected version, November 1995 \end{flushright}
\vspace{2cm}
%---------------------titre ---------------------------------------

\begin{center}
{\Large {\bf Tests of Anomalous Higgs Boson Couplings through $e^-e^+ \to ZH$
and $\gamma H$}} \footnote{%
Partially supported by the EC contract CHRX-CT94-0579.} \vspace{1.5cm} \\%
{\large G.J. Gounaris$^{a}$, F.M. Renard$^b$ and N.D. Vlachos$^{a}$} \vspace
{0.5cm} \\

$^a$Department of Theoretical Physics, University of Thessaloniki,\\%
Gr-54006, Thessaloniki, Greece,\\\vspace{0.2cm} $^b$Physique
Math\'{e}matique et Th\'{e}orique, CNRS-URA 768,\\Universit\'{e} de
Montpellier II,  F-34095 Montpellier Cedex 5.\\

\vspace {2.0cm}

{\bf Abstract}
\end{center}

\noindent
We show how the processes $e^- e^+ \to ZH, ~ \gamma H$, which will be
studied at LEP2 and at high energy colliders, could be used to search for
types of New Physics (NP) characterized by an effective NP scale in the few
TeV range and affecting the scalar sector only. In particular, for $e^- e^+
\to Z H$, we propose the measurement of suitable azimuthal asymmetries
determined from the angular distribution of the $Z$-decay plane with respect
to the $Z$-production one. Such asymmetries, together with the $ZH$ angular
distribution, may allow a complete disentangling of the five $dim=6$
operators $ {\cal O }_{UB}$, $ {\cal O }_{UW}$, $\overline{{\cal O }}_{UB}$,
$\overline{{\cal O }}_{UW}$, ${\cal O }_{\Phi2}$, describing the residual
NP effects to the Higgs couplings. We note that the first four of these
operators contribute to processes relevant to LEP1 precision measurements
only at the 1-loop level, while the last operator remains completely
unconstrained at this level. The process $e^- e^+ \to \gamma H$ is also very
sensitive to NP and should bring important independent information.

\clearpage

\section{Introduction}

The scalar sector is the most mysterious part of the Standard Model (SM) of
the electroweak interactions and the favorite place for generating New
Physics (NP) manifestations \cite{review, Djouadi}. If it happens that all
new particles are too heavy to be directly produced at future colliders,
then the only way NP could manifest itself, is through residual interactions
affecting the particles already present in SM. The possibility for such
residual NP effects involving the interactions of the gauge bosons with the
leptons and the light quarks \cite{deR, Hag}, has already been essentially
excluded by LEP1 \cite{Schaile}. Thus, only in self-interactions among the
gauge and Higgs bosons \cite{Hag1} and possibly also the heavy quarks \cite
{GRVZbb}, there appears to be still room for some NP.

In the following, we will concentrate on types of NP affecting the purely
weak boson sector only. Those involving the gauge bosons only (like
{\it e.g.\/} the $\gamma WW$ and $ZWW$ couplings) will be
constrained by $e^{-}e^{+}\to W^{-}W^{+}$ at LEP2 \cite{Gaemers,
Bilenky} and NLC \cite{BMT}, as well as by other gauge boson
pair production processes \cite{bb, phys, Boudj}.
Assuming that this has been done and nothing new has been found,
then the only remaining possibility for NP in the purely bosonic sector will
be hidden in interactions involving the physical Higgs boson. This assumes
of course that the Higgs particle exists and has a mass comparable to $M_W$.
Supposing that the NP scale satisfies $\Lambda _{NP}\gg M_W$ and that it is
justified to restrict to $SU(2)\times U(1)$ gauge invariant $dim=6$
operators only, the NP should then be determined by the four ''blind''
operators ${\cal O}_{UB}$, ${\cal O}_{UW}$ and their
$CP$-violating partners $\overline{{\cal O}}_{UB}$,
$\overline{{\cal O}}_{UW}$, as well as the
''superblind'' operator ${\cal O}_{\Phi 2}$ \cite{Hag, Hag1, dyn1}. This
possibility is also motivated by a recent treatment suggesting that it is
much easier to create anomalous NP interactions for the Higgs particle  than
for the gauge bosons \cite{dyn2, dyn1}. The basic reason for this is that
Higgs couplings enjoy the familiar Yukawa type freedom, while the gauge
boson interactions are strongly constrained by the gauge principle.

It is important to state that the operators ${\cal O}_{UB}$, ${\cal O}_{UW}$,
$\overline{{\cal O}}_{UB}$, $\overline{{\cal O}}_{UW}$, are only mildly
constrained (at the 1-loop level) by LEP1 precision measurements, as opposed
to a number of other $dim=6$ operators which are strongly constrained
through their non-vanishing tree level contributions. In a sense, these
operators cannot ''see'' LEP1 and they have thus been called ''blind'' \cite
{deR, Hag}. In that respect, the operator ${\cal O}_{\Phi 2}$ is
''superblind'', since it is insensitive to LEP1 measurements even at 1-loop.
It turns out that at tree level and up to first order in the anomalous
couplings, $\O_{\Phi2}$  is also insensitive
to the $\gamma \gamma $ collider searches studied in
\cite{higpro1, higpro2, ggVV1,
ggVV2}. In this ~sense ${\cal O}_{\Phi 2}$ is ''blind'' to all
these experiments and it can
''see'' an $e^{-}e^{+}$ collider only through the process $e^{-}e^{+}\to ZH$
\cite{Hag1, dyn1}.

Of course, the study of all these operators requires Higgs boson exchange or
production processes. Such a study is relevant only in
cases where the
Higgs particle is sufficiently light and can therefore  be produced at LEP2
or NLC. This is certainly possible if the NP scale is at
the $TeV$ range \cite{Isidori}.

The subject of this paper is to study these NP effects on the
processes $e^{-}e^{+}\to ZH,\,\gamma H$. In SM, the process
$e^{-}e^{+}\to Z\to ZH$ is
allowed at tree level (provided $m_H<\sqrt{s}-M_Z$) \cite{Bj},
while $e^{-}e^{+}\to \gamma H$ is only possible at 1-loop level
(if $m_H<\sqrt{s}$) \cite{Barro, HgAA}.
We use the same general framework as in \cite{Hag1,
Barger}. Here, in addition, we study in more detail the angular distribution
in $e^{-}e^{+}\to Z_{f\bar f}H$ and include the standard
contribution to $e^{-}e^{+}\to \gamma H$.
For other recent works on these processes see \cite{HagCP, Ma}.
The complementary possibilities offered at higher energies for
the first four of the above operators, where laser induced $\gamma \gamma $
collisions may be used to study the production of a single $H$ or a boson
pair, have been described in \cite{ggVV1, ggVV2, higpro1, higpro2}.

The precision tests for $e^{-}e^{+}\to ZH$ consist of measuring the angular
distribution $d\sigma /d\cos \vartheta $ and the $Z$ helicity density matrix
elements through the analysis of the $Z\to f\bar f$ decay distribution;
especially its dependence on the azimuthal angle $\phi _f$.
Here, $\vartheta $
is defined as the angle between the $e^{-}$ beam and the $Z$ direction in
the $e^{-}e^{+}$ c.m. frame, while $\phi _f$ is the azimuthal angle of the
charged fermion $f$ with respect to the $ZH$ production plane. In order to
disentangle the contributions generated from the five operators
above, we need
suitable independent observables. We show that in addition to the $Z$
angular distribution $d\sigma /d\cos \vartheta $, four azimuthal asymmetries
determined respectively by the coefficients of the $\cos \phi _f$, $\sin
\phi _f$, $\sin 2\phi _f$ and $\cos 2\phi _f$ terms of the $f\bar f$ plane
distribution, will be ~useful for this search.

If the NP scale $\Lambda _{NP}$ lies in the $TeV$ range, the number of events
at LEP2 will be too small to allow for such a detailed analysis. Thus, at
LEP2 only $d\sigma /d\cos \vartheta $ will be measurable and will give a
meaningful constraint to a combination of NP couplings (essentially the
anomalous $HZZ$ coupling). At NLC this study can be further pursued and a
complete disentangling of the five couplings could then be expected,
{}~particularly if the results from $e^{-}e^{+}\to ZH$ are combined with those
from the processes studied in \cite{higpro1, higpro2, ggVV1, ggVV2}.

The differential cross section for $e^{-}e^{+}\to \gamma H$ provides also a
very sensitive test of NP, in situations where the final photons are easily
detected. In particular, if it happens that $m_H>100GeV$, this process may
be the only way to produce $H$ at LEP2. The operators studied ~through this
process are ${\cal O}_{UB}$, ${\cal O}_{UW}$, $\overline{{\cal
O}}_{UB}$, $\overline{{\cal O}}_{UW}$. Here, it is impossible to
disentangle the
$CP$-concerving pair (${\cal O}_{UB}$, ${\cal O}_{UW}$) from the
$CP$-violating
one ($\overline{{\cal O}}_{UB}$, $\overline{{\cal O}}_{UW}$). Note ~also
that the combination of the ${\cal O}_{UB}$ and ${\cal O}_{UW}$ couplings
appearing in the $\gamma H$ angular distribution is different from the one
appearing in $ZH$. Thus, the $e^{-}e^{+}\to \gamma H$ ~differential cross
section can be used to help disentangle ${\cal O}_{UB}$ from ${\cal O}_{UW}$,
without going through the difficult analysis of the $Z$ spin density matrix.
This later analysis would still be needed though to disentangle the
$CP$-violating effects induced by $\overline{{\cal O}}_{UB}$ and $\overline{%
{\cal O}}_{UW}$.

In Section 2, we compute the helicity amplitudes for both $e^{-}e^{+}\to ZH$
and $e^{-}e^{+}\to \gamma H$ processes due to SM and NP contributions. The
corresponding cross sections together with the other observables needed to
discriminate among the effects of the five operators are
discussed in Section 3
while the Z density matrix elements are derived in Appendices A, B. The
results for the LEP2 and NLC conditions are presented and commented in the
last Section 4, while in Section 5 we give our conclusions.

\section{Anomalous Higgs couplings and $e^-e^+\to ZH$, $\gamma H$ amplitudes}

As discussed in the introduction and in previous papers \cite{Hag, Hag1,
ggVV1, higpro1}, there are only five $SU(2)\times U(1)$ gauge
invariant $dim=6$ operators relevant to $e^{-}e^{+}\to ZH$,
$\gamma H$. These consist
of the four blind operators
\begin{eqnarray}
{\cal O}_{UW} &=&\frac 1{v^2}\,(\Phi ^{\dagger }\Phi -\frac{v^2}2)\,%
\overrightarrow{W}^{\mu \nu }\cdot \overrightarrow{W}_{\mu \nu }\ \ \ ,\ \
\\[0.1cm]
{\cal O}_{UB} &=&\frac 4{v^2}~(\Phi ^{\dagger }\Phi -\frac{v^2}2)B^{\mu \nu
}\ B_{\mu \nu }\ \ \ ,\ \  \\[0.1cm]
\overline{{\cal O}}_{UW} &=&\frac 1{v^2}\,(\Phi ^{\dagger }\Phi )\,%
\overrightarrow{W}^{\mu \nu }\cdot \widetilde{\overrightarrow{W}}_{\mu \nu
}\ \ \ ,\ \  \\[0.1cm]
\overline{{\cal O}}_{UB} &=&\frac 4{v^2}~(\Phi ^{\dagger }\Phi )B^{\mu \nu
}\ \widetilde{B}_{\mu \nu }\ \ \ ,\ \ \
\end{eqnarray}
and the ''superblind'' one\footnote{%
Actually there exists also another superblind operator called $ {\cal O }%
_{\Phi3}=8 (\Phi^\dagger \Phi)^3$ which only gives an irrelevant Higgs mass
renormalization and can therefore be neglected.}
\begin{equation}
{\cal O}_{\Phi 2}~=~4\partial _\mu (\Phi ^{\dagger }\Phi )\partial ^\mu
(\Phi ^{\dagger }\Phi )\ \ \ \ ,\
\end{equation}
which is insensitive to ~LEP1 physics at the 1-loop level, but sensitive to
the process $e^{-}e^{+}\to ZH$ \cite{Hag, Hag1, dyn1}. As usual, the vacuum
{}~expectation value of the Higgs field is denoted by
$v=(G_F\sqrt{2})^{-1/2}$. The NP effective lagrangian
determining the couplings of these operators
is given by
\begin{equation}
{\cal L}_{NP}=d\ {\cal O}_{UW}+{\frac{d_B}4}\ {\cal O}_{UB}+\overline{d}\
\overline{{\cal O}}_{UW}+{\frac{\overline{d}_B}4}\
\overline{O}_{UB}+ \frac{f_{\Phi 2}}{v^2}{\cal O}_{\Phi 2}\ \ \ \ \ \ .\
\end{equation}

The only effect of ${\cal O}_{\Phi 2}$ at the tree level is to induce a wave
function renormalization to the Higgs field given by
\begin{equation}
Z_H~=~\frac 1{1+8f_{\Phi 2}}~\simeq ~1-8f_{\Phi 2}\ \ \ .\
\end{equation}
Here, as well as in the ~previous works \cite{higpro1, higpro2,
ggVV1, ggVV2}, we only consider tree level anomalous
contributions and we restrict to
cases where only one of the operators above acts each time.
Thus, the ${\cal O}_{\Phi 2}$ contribution can only be studied
in $e^{-}e^{+}\to ZH$.
In particular, in this framework there is no ${\cal O}_{\Phi 2}$
contribution to
$e^{-}e^{+}\to \gamma H$ or to the $\gamma \gamma $ and $e\gamma $-Collider
processes considered in \cite{higpro1, higpro2, ggVV1, ggVV2}. Contributions
from ${\cal O}_{\Phi 2}$ to these later processes could arise at  tree level
only if terms bilinear in the anomalous couplings of two different operators
were to be retained.

In the following, it will be convenient to introduce the definitions
\begin{equation}
d_{\gamma Z}=s_Wc_W(d-d_B)\ \ ,\ d_{ZZ}=dc_W^2+d_Bs_W^2\ \ ,\ d_{\gamma
\gamma }=ds_W^2+d_Bc_W^2\ ,
\end{equation}
as well as the corresponding $CP$-violating ones for the
$\ol{d},~\ol{d}_B$ ~couplings defined
in (6).

The invariant amplitude for $e_\lambda ^{-}e_{-\lambda }^{+}\to
Z_\tau H$ takes then the simple compact form (see also \cite{HagCP})
\begin{eqnarray}
\null  &\null &F_{\lambda \tau }(e^{-}e^{+}\to ZH)~=~  \nonumber \\
\null \null  &&{\frac{\lambda eg\sqrt{Z_H}}{M_W\sqrt{s}}}\Bigg ({\frac{M_Z^2%
}{4c_Ws_W}}\chi (\lambda A_e-V_e)\Bigg [(1-{\tau }^2)~{\frac{k_0\sin
\vartheta }{M_Z}}-{\tau }^2~{\frac{\tau cos\vartheta +\lambda }{\sqrt{2}}}%
\Bigg ]  \nonumber \\
\null + &&[d_{\gamma Z}+d_{ZZ}\,{\frac \chi {4s_Wc_W}}(\lambda
A_e-V_e)][(1-\tau ^2)~2M_Z\sqrt{s}\,\sin \vartheta -\tau ^2k_0\sqrt{2s}(\tau
\cos \vartheta +\lambda )]  \nonumber \\
\null + &&i[\overline{d}_{\gamma Z}+\overline{d}_{ZZ}\,{\frac \chi {4s_Wc_W}}%
(\lambda A_e-V_e)][\tau ^2k\sqrt{2s}(\cos \vartheta +\lambda \tau )]\Bigg) \
,\
\end{eqnarray}
where $\tau $ is the $Z$ helicity and $\lambda =\pm 1$ is the difference
between the $e^{-}$ and $e^{+}$ helicities. Moreover\footnote{
Of course close to the $Z$-pole we should replace
$\mz \to \mz -i\Gamma_Z/2$.}, $\chi =s/(s-M_Z^2)$
while $V_f=2t_f^{(3)}-4Q_fs_W^2$, $A_f=2t_f^{(3)}$ are the standard $Zf\bar f
$ vector and axial couplings, with $t_f^{(3)}$ being the third component of
the ~left-isospin of the fermion$\ f$ . The quantities $k$,
$k_0$ and $\vartheta $ are the momentum, energy and production
angle of $Z$ in the $e^{-}e^{+}$ c.m. frame respectively,
with the z-axis defined by the $e^{-}$
direction. The normalization of the invariant amplitude is such that
\begin{equation}
\frac{d\sigma (e^{-}e^{+}\to ZH)}{d\cos \vartheta }~=~\frac k{64\pi
s^{3/2}}\sum_{\lambda ,\tau }|F_{\lambda \tau }|^2\ .
\end{equation}

The first term in (9) gives the SM result, (apart of course from
the ${\cal O}_{\Phi 2}$ contribution determined by the overall
factor $\sqrt{Z_H}$). It
involves the production of both transverse $Z_T$ and longitudinal $Z_L$, but
it is dominated by $Z_L$ at high energies. The second term
determines the $CP$-conserving anomalous contribution
from ${\cal O}_{UB}$ and
${\cal O}_{UW}$. It again involves production of both $Z_T$ and
$Z_L$, but  it is $Z_T$ now
that ~dominates at high energies. The third term gives the $CP$-violating
contribution which is purely $Z_T$ and grows with the energy
like the $Z_T$ $CP$-conserving one. Concerning the overall
factor $\sqrt{Z_H}$ in (9), we
remark that according to the ~approximations explained above where only
linear terms in the anomalous couplings were ~considered, we must
substitute $Z_H\to 1$ whenever contributions from the last two
terms in (9) are
evaluated. We  note also, that the $CP$-conserving ($CP$-violating) part of
the amplitude in (9) satisfies
\begin{eqnarray}
F_{++}(\cos \vartheta ) &=&\pm F_{+-}(-\cos \vartheta )\ \ , \\
F_{-+}(\cos \vartheta ) &=&\pm F_{--}(-\cos \vartheta )\ \ ,\
\end{eqnarray}
where the upper sign is for the $CP$-conserving part and the lower sign for
the $CP$-violating one. These relations mean that $d\sigma /d\cos \vartheta $
for unpolarized $e^{\pm }$ beams is symmetric under $\cos \vartheta \to
-\cos \vartheta $; compare (19) below.

We next turn to $e_\lambda ^{-}e_{-\lambda }^{+}\to \gamma _\tau H$ for
which we obtain
%\newpage
\begin{eqnarray}
&\null &F_{\lambda \tau }^{(A)}(e^{-}e^{+}\to \gamma H)={\frac{\lambda ~2%
\sqrt{2}~\pi \alpha }{s_WM_W\sqrt{s}}}\,(s-m_H^2)\cdot   \nonumber \\
&&\null \Bigg \{ -(\tau \cos \vartheta +\lambda )[d_{\gamma \gamma
}+d_{\gamma Z}{\frac \chi {4s_Wc_W}}(\lambda A_e-V_e)]  \nonumber \\
&&\null +~i(\cos \vartheta +\tau \lambda )[\overline{d}_{\gamma \gamma }+%
\overline{d}_{\gamma Z}{\frac \chi {4s_Wc_W}}(\lambda A_e-V_e)]\Bigg \}\ \
,\
\end{eqnarray}
\begin{eqnarray}
&\null &F_{\lambda \tau }^{(SM)}(e^{-}e^{+}\to \gamma H)={\frac{\lambda
\alpha ^2\sqrt{2}}{M_Ws_W\sqrt{s}}}\,(s-m_H^2)\cdot   \nonumber \\
&&\null \Bigg \{ (\tau \cos \vartheta +\lambda )[I_{\Delta G}+\frac \chi
{4s_W^2}(V_e-\lambda A_e)I_{\Delta Z}]  \nonumber \\
&&\null +~\frac{s(1-\lambda )}{32s_W^2M_W^2}[(\tau +\lambda )(1+\cos
\vartheta )I_W^{+}(\vartheta )+(\lambda -\tau )(1-\cos \vartheta
)I_W^{-}(\vartheta )]  \nonumber \\
&&\null +~\frac s{128s_W^2M_W^2}(V_e^2+A_e^2-2\lambda V_eA_e)[(\tau +\lambda
)(1+\cos \vartheta )I_Z^{+}(\vartheta )  \nonumber \\
&&\null +(\lambda -\tau )(1-\cos \vartheta )I_Z^{-}(\vartheta )]\Bigg \} \ \
\ ,\
\end{eqnarray}
for the anomalous and the 1-loop SM contributions respectively. Here, $%
\vartheta $ is the c.m. angle of the final $\gamma $ with respect to the $%
e^{-}$ axis. The normalization of the amplitudes is the same as
in the case of $e^{-}e^{+}\to ZH$. In writing (13,14)
we have again neglected
tree-level contributions quadratic in the anomalous couplings, and also
1-loop contributions linear in the anomalous couplings. This implies that no
contribution from ${\cal O}_{\Phi 2}$ should be included and thus no $\sqrt{%
Z_H}$ appears in (13,14).

The SM contribution to $e_\lambda ^{-}e_{-\lambda }^{+}\to \gamma _\tau H$
is given by\footnote{%
A new analysis of the SM contribution has just appeared in \cite{HgAA}. The
computation is done in a non linear gauge and the results are expressed
in terms of a rather large number of dilogarithm functions. The
numerical results for $\sigma(e^-e^+ \to \gamma H)$ are consistent with
those of \cite{Barro} but a little larger for $s \neq M_Z^2$, which gives a
measure of the accuracy of the methods eventually used. This is
not a problem for our purpose though, because the
predicted SM rate seems unobservable in any case.} \cite{Barro} in terms of
the complex functions $I_{\Delta Z}(s)$, $I_{\Delta G}(s)$ describing the $Z$
and $\gamma $ exchange diagrams, and the functions $I_W^{\pm }(s,\vartheta )$%
, $I_Z^{\pm }(s,\vartheta )$ describing the $W$ and $Z$ boxes respectively.
These functions are expressed by \cite{Barro} as double integrals in the
Feynman-parameter space\footnote{See Eqs. (2.4a,4b,21,22,26,27),
(A.3,4,8) in \cite{Barro}.}. The most difficult ones to calculate
are $I_{\Delta Z}$, $%
I_{\Delta G}$, $I_W^{\pm }$ , for which there exist singularity lines within
the integration region for LEP2 energies and beyond.
 The numerical integration around these singularities
can be done by using standard routines and  Feynman's $i\epsilon $
prescription in order to separate principal-value and absorptive
contributions. Such a separation should always be possible provided that
these integrals are selected so that they can accept a dispersive
Mandelstam-type representation like the one satisfied by any Feynman
amplitude.

We also note that \cite{Barro}
\begin{equation}
I_W^{-}(s,\vartheta )~=~I_W^{+}(s,\pi -\vartheta )\ \ ,\
\end{equation}
while
\begin{equation}
I_Z^{-}(s,\vartheta )~=~-\,I_Z^{+}(s,\pi -\vartheta )\ \ .\
\end{equation}
It then turns out that for LEP2 and NLC energies the $Z$-box contribution is
rather small compared to the other ones, a result
which is related to the fact
that the relative integral has no absorptive part. This implies that the SM
as well as the anomalous contribution to the amplitude for $e_\lambda
^{-}e_{-\lambda }^{+}\to \gamma _\tau H$ satisfy approximately (11,12),
which in turn leads to the conclusion that
$d\sigma (e^{-}e^{+}\to \gamma H)/d\cos \vartheta $
for unpolarized beams is essentially forward-backward symmetric.

\section{Observables for $e^-e^+\to Z_{(f\bar f)} H$ and $e^-e^+\to \gamma H$%
.}

The $Z$ spin density matrix in the helicity basis $\rho _{\tau \tau ^{\prime
}}$ is calculated from the amplitude in (9) and given in Appendix A.
Using that and ~assuming that the decay $Z\to f\bar f$ is standard, the
angular distribution for the $f\bar f$ system in the $Z$ rest frame is given
by
\begin{equation}
{\frac{d\sigma (e^{-}e^{+}\to Z_{f\bar f}H)}{d\cos \vartheta d\Omega _f}}={%
\frac{kB(Z\to f\bar f)}{64\pi s\sqrt{s}}}\sum_{\tau \tau ^{\prime }}\rho
_{\tau \tau ^{\prime }}Q_{\tau \tau ^{\prime }}\ \ ,\
\end{equation}
where $\Omega \equiv (\theta _f,\phi _f)$ are the angles of the fermion$\ f$
in the $Z$-rest frame with the z-axis chosen along the $Z$ momentum
{$\bf k$} while the y-axis is along {$\bf e^{-}\times k$}. The
quantities $Q_{\tau \tau ^{\prime }}$ are given in Appendix B and $B(Z\to
f\bar f)$ is the $Z$ branching ratio to $f\bar f$.

The first observable from $e^-e^+\to ZH $ is the $Z$ differential cross
section obtained by integrating over the $f$-fermion solid angle
\begin{equation}
{\frac{d \sigma(e^-e^+\to ZH )}{d \cos\vartheta}} = {\frac{1}{B(Z\to f \bar
f)}}\int d\Omega_f {\frac{d \sigma(e^+e^-\to Hf\bar f) }{d \cos\theta d
\Omega_f}} \ \
\end{equation}
which for unpolarized beams gives the explicit formula
\begin{eqnarray}
&& {\frac{d\sigma(e^-e^+ \to ZH)}{d \cos\vartheta}} = {\frac{k}{64\pi s\sqrt{%
s}}}(\rho_{++}+\rho_{--}+\rho_{--})  \nonumber \\
&=&{\frac{\alpha^2 \pi k M^2_Z}{4 s^2_W c^2_W s^{5/2}}}Z_H\, \Bigg \{ {\frac{%
(1-4s^2_W+8s^4_W) }{4s^2_W}}\chi^2~ \left [\frac{2}{c^2_W}+{\frac{k^2}{M_W^2}%
} \sin^2\vartheta \right ]  \nonumber \\
&-&{\frac{4 \chi k_0 \sqrt{s} }{s_W c_W M_Z^2 }}~ [2s^2_W D_+-(1-2s^2_W)D_-]
\nonumber \\
&+&{\frac{4s }{M^2_Z}}(D^2_+ + D^2_-)[2+(1+\cos^2\vartheta){\frac{k^2}{M^2_Z}%
}]  \nonumber \\
&+&{\frac{4s k^2 }{M^4_Z}}(1+\cos^2\vartheta)(\overline{D}^2_++\overline{D}%
^2_-) \Bigg \} \ ,\
\end{eqnarray}
where $D_+,~ D_-$ are given in (A.2,A.3) and correspondingly for the
$CP$-violating couplings $\overline{D}_+,~ \overline{D}_-$. As already noted,
this expression is symmetric under the interchange $\vartheta \to \pi
-\vartheta$.

An additional set of interesting four observables depending only on the four
operators ${\cal O}_{UB}$, ${\cal O}_{UW}$, $\overline{{\cal O}}_{UB}$, $%
\overline{{\cal O}}_{UW}$ may be constructed by integrating (17) over $\cos
\theta _f$ and the $Z$ production angle $\vartheta $. This gives
\begin{eqnarray}
{\frac{d\sigma (e^{-}e^{+}\to Z_{f\bar f}H)}{d\phi _f}} &\sim
&\int_{-1}^1d\cos \vartheta \int_{-1}^1d\cos \theta _f\sum_{\tau \tau
^{\prime }}\rho _{\tau \tau ^{\prime }}Q_{\tau \tau ^{\prime }}  \nonumber \\
&\sim &[B+b_{13}\sin 2\phi _f+b_8\cos 2\phi _f+b_{12}\sin \phi _f+b_{14}\cos
\phi _f]\ ,\
\end{eqnarray}
where $b_{13}$, $b_8$, $b_{12}$ and $b_{14}$ depend only on $s$, and the
four anomalous $d$-couplings defined in (6). The $f_{\Phi 2}$ coupling of $%
{\cal O}_{\Phi 2}$ is completely factored out and does not appear in these
coefficients.

The $b$-quantities in (20) may be obtained by constructing suitable
{}~azimuthal asymmetries. To this end we integrate the above differential
cross section with respect to $\phi _f$ over the regions $[0,\pi /4]$, $[\pi
/4,\pi /2]$, $[\pi /2,3\pi /4]$, $[3\pi /4,\pi ]$, $[\pi ,5\pi /4]$, $[5\pi
/4,3\pi /2]$, $[3\pi /2,7\pi /4]$, $[7\pi /4,2\pi ]$, calling the respective
cross sections $\sigma _{1a}$, $\sigma _{1b}$, $\sigma _{2a}$, $\sigma _{2b}$%
, $\sigma _{3a}$, $\sigma _{3b}$, $\sigma _{4a}$, $\sigma _{4b}$, where the
number in the index counts the quadrant in which $\phi _f$ lies. This way,
we define the asymmetries
\begin{eqnarray}
A_{13} &=&\frac{\sigma _{1a}+\sigma _{1b}+\sigma _{3a}+\sigma _{3b}-\sigma
_{2a}-\sigma _{2b}-\sigma _{4a}-\sigma _{4b}}\sigma ~=~\frac{2b_{13}}{\pi B}%
\ \ , \\
A_{12} &=&\frac{\sigma _{1a}+\sigma _{1b}+\sigma _{2a}+\sigma _{2b}-\sigma
_{3a}-\sigma _{3b}-\sigma _{4a}-\sigma _{4b}}\sigma ~=~\frac{2b_{12}}{\pi B}%
\ \ , \\
A_{14} &=&\frac{\sigma _{1a}+\sigma _{1b}+\sigma _{4a}+\sigma _{4b}-\sigma
_{2a}-\sigma _{2b}-\sigma _{3a}-\sigma _{3b}}\sigma ~=~\frac{2b_{14}}{\pi B}%
\ \ , \\
A_8 &=&\frac{\sigma _{1a}+\sigma _{2b}+\sigma _{3a}+\sigma _{4b}-\sigma
_{1b}-\sigma _{2a}-\sigma _{3b}-\sigma _{4a}}\sigma ~=~\frac{2b_8}{\pi B}\ \
,
\end{eqnarray}
where
\begin{eqnarray}
B &=&{\frac{M_Z^2\chi ^2(1-4s_W^2+8s_W^4)}{4s_W^2c_W^2}}[2M_Z^2+k_0^2]-\frac{%
6M_Z^2k_0\sqrt{s}\chi }{s_Wc_W}[2s_W^2D_{+}-(1-2s_W^2)D_{-}]  \nonumber \\
\null \null  &&+\,4s(M_Z^2+2k_0^2)(D_{+}^2+D_{-}^2)+8sk^2(\overline{D}_{+}^2+%
\overline{D}_{-}^2)\ \ , \\[0.1cm]
b_{13} &=&-\frac{k\sqrt{s}M_Z^2\chi }{s_Wc_W}(2s_W^2\overline{D}%
_{+}-(1-2s_W^2)\overline{D}_{-})+4kk_0s(D_{+}\overline{D}_{+}+D_{-}\overline{%
D}_{-})\ \ , \\
b_{12} &=&-\,{\frac{9\pi ^2M_Z}{16}}\left( \frac{V_fA_f}{V_f^2+A_f^2}\right) %
\Bigg [ 4sk(D_{-}\overline{D}_{-}-D_{+}\overline{D}_{+})  \nonumber \\
\null \null  &&+\,\frac{\chi kk_0\sqrt{s}}{s_Wc_W}[2s_W^2\overline{D}%
_{+}+(1-2s_W^2)\overline{D}_{-}]\Bigg ] \ \ , \\
b_{14} &=&-\,{\frac{9\pi ^2M_Z}{16}}\left( \frac{V_fA_f}{V_f^2+A_f^2}\right) %
\Bigg [\frac{M_Z^2(1-4s_W^2)\chi ^2k_0}{4s_W^2c_W^2}  \nonumber \\
\null \null  &&+\,{\frac{\sqrt{s}\chi }{s_Wc_W}}%
(k_0^2+M_Z^2)[2s_W^2D_{+}+(1-2s_W^2)D_{-}]+4sk_0(D_{-}^2-D_{+}^2)\Bigg ] \ \
,\
\end{eqnarray}
\newpage
\begin{eqnarray}
b_8 &=&{\frac{M_Z^4(1-4s_W^2+8s_W^4)}{8s_W^2c_W^2}}~\chi ^2-~{\frac{%
M_Z^2\chi k^0\sqrt{s}}{s_Wc_W}}(2s_W^2D_{+}-(1-2s_W^2)D_{-})  \nonumber \\
\null \null  &&+~2sk_0^2(D_{+}^2+D_{-}^2)-2sk^2(\overline{D}_{+}^2+\overline{%
D}_{-}^2)]\ \ .\
\end{eqnarray}
The  above way for measuring these asymmetries assumes of course that the
whole $\phi _f$ range is covered. ~Otherwise, these asymmetries should be
{}~measured by fitting the $\phi _f$ distribution using (20).

Turning now to the differential cross section for
$e^{-}e^{+}\to \gamma H$, assuming unpolarized beams
and neglecting the SM contribution we get
\begin{eqnarray}
\frac{d\sigma (e^{-}e^{+}\to \gamma H)}{d\cos \vartheta } &=&\frac{\pi
\alpha ^2}{8M_W^2s_W^2}\left( 1-\frac{m_H^2}s\right) ^3(1+\cos ^2\vartheta
)\cdot   \nonumber \\
\null \null  &&\Bigg \{ (d^2+\overline{d}^2)\left[ 2s_W^4+\chi
(1-4s_W^2)s_W^2+\chi ^2(\frac 14-s_W^2+2s_W^4)\right]   \nonumber \\
\null + &&(d_B^2+\overline{d}_B^2)\left[ 2c_W^4-\chi (1-4s_W^2)c_W^2+\chi
^2(\frac 14-s_W^2+2s_W^4)\right]   \nonumber \\
\null + &&(dd_B+\overline{d}\overline{d}_B)\Big [4s_W^2c_W^2+\chi
(1-4s_W^2)(1-2s_W^2)  \nonumber \\
\null \null  &&-2\chi ^2(\frac 14-s_W^2+2s_W^4)\Big ] \Bigg \} \ \ .\
\end{eqnarray}
We note though that in the numerical calculations presented
below we have used the complete expressions given by the
amplitudes in (13,14) which include both the SM and the
anomalous contributions.

\section{Analysis at LEP2 and NLC.}

Before discussing the prospects for detecting NP at LEP2 and NLC, let us
first describe what we can reasonably expect for the magnitude of the
anomalous couplings appearing in the effective NP Lagrangian in (6). The
most straightforward way to estimate the possible magnitude of these
coupling is by using the unitarity relations. For the operators $ {\cal O }%
_{UB}$, $ {\cal O }_{UW}$, $\overline{{\cal O }}_{UB}$, $\overline{{\cal O }}%
_{UW}$, $ {\cal O }_{\Phi2}$ these relations are already
known \cite{uni1,uni2, dyn2}. Calling
$\Lambda_{NP}$ the operator dependent NP
scale where the related forces become strong, these
unitarity relations may be written
as\footnote{These unitarity relations are obtained
by making simple fits to the expression
for the maximum eigenvalue of the partial
wave transition matrix. For the operators in (1)-(4), the fits
in \cite{uni1,uni2, dyn2} were occasionally valid only up to
$3 TeV$. Here we give better ones valid for any
$\Lambda_{NP}\gsim 1TeV$.} \cite{ggVV2}
\begin{eqnarray}
d & \simeq & \frac{104.5~\left ({\frac{M_W}{\Lambda_{NP}}}
\right )^2} {1+6.5 \left ({\frac{M_W}{\Lambda_{NP}}}\right )} \ ,
\mbox{ for } d>0 \ , \nonumber \\
d & \simeq & -~ \frac{104.5~\left ({\frac{M_W}{\Lambda_{NP}}}
\right )^2} {1- 4 \left ({\frac{M_W}{\Lambda_{NP}}}\right )} \ ,
\mbox{ for }\  d<0 \ , \  \\
d_B & \simeq & \frac{195.8 ~\left ({\frac{M_W}{\Lambda_{NP}}}
\right )^2} {1+200 \left  ({\frac{M_W}{\Lambda_{NP}}}
\right )^2}\ ,
\mbox{ for } d_B>0 \ , \nonumber \\
d_B & \simeq & -~ \frac{195.8 ~\left ({\frac{M_W}{\Lambda_{NP}}}
\right )^2} {1 +50 \left  ({\frac{M_W}{\Lambda_{NP}}}
\right )^2}\ ,
\mbox{ for }\  d_B<0 \ , \  \\
|\ol{d}| & \simeq & \frac{104.5~\left ({\frac{M_W}{\Lambda_{NP}}}
\right )^2} {1+3 \left ({\frac{M_W}{\Lambda_{NP}}}\right )} \ ,
\ \\
|\ol{d}_B| & \simeq & \frac{195.8 ~\left ({\frac{M_W}{\Lambda_{NP}}}
\right )^2} {1+ 100 \left  ({\frac{M_W}{\Lambda_{NP}}}
\right )^2}\  \  . \
\end{eqnarray}
Applying (31-34) for $\Lambda_{NP}=1 TeV$,
we get $d
\simeq 0.4 $ or $-1.$, $d_B \simeq
0.6 $ or $-1.$, $|\overline{d}| \simeq 0.5 $, $|\overline{d}_B|
\simeq 0.8 $ which
give an estimate of the possible ~largest values for these couplings.
Similar indications are also obtained from an 1-loop analysis of the LEP1
precision measurements \cite{Hag, Hag1}.

For the superblind operator $ {\cal O }_{\Phi2}$, a similar as in \cite
{uni1, uni2, dyn2} unitarity analysis derives its strongest
result from the $J=0$ (partial wave) transition amplitudes
involving the channels $|HH>$, $|W^-W^+~ LL>$, $|ZZ~ LL>$.
Such a study gives unitarity saturation for
\begin{equation}
\Bigg |\frac{f_{\Phi2}}{(1+8\,f_{\Phi2})^2} \Bigg |~
\simeq ~ \frac{1}{a_w} \ \ ,
\end{equation}
where
\begin{equation}
a_w ~=~ \frac{\alpha \sqrt 3}{4 s^2_W}~ \frac{\Lambda^2_{NP}}{M_W^2} \ \ . \
\end{equation}
Solving (35), we obtain that for $\Lambda_{NP}<3.7 TeV$ ({\it
i.e.\/}\@ $0< a_w < 32$) there is no unitarity constrain for $f_{\Phi2}>0$;
while for $f_{\Phi2} <0$ we get
\begin{eqnarray}
f_{\Phi2} & \simeq & \frac{-16 -a_w+\sqrt{a_w (a_w +32)}}{128} \ \ \ \ \ \
.\
\end{eqnarray}
On the contrary, for $\Lambda_{NP}>3.7 TeV$ ({\it i.e.\/}\@ $a_w>32$) we get
\begin{eqnarray}
f_{\Phi2}& \simeq & \frac{-16 +a_w-\sqrt{a_w (a_w -32)}}{128} \ \ \ %
\makebox{ for } f_{\Phi2} >0 \ \ \ ,  \nonumber \\
f_{\Phi2} & \simeq & \frac{-16 -a_w+\sqrt{a_w (a_w +32)}}{128} \ \ \ %
\makebox{ for } f_{\Phi2} <0 \ \ \ , \
\end{eqnarray}
which for $\Lambda_{NP}\gg 3.7 TeV $ gives
\begin{eqnarray}
|f_{\Phi2}| & \simeq & 75\frac{M^2_W}{\Lambda^2_{NP}} \ \ .
\end{eqnarray}
Thus, if \eg\@ $|f_{\Phi 2}| \simeq 0.004$, then the
$ {\cal O }_{\Phi2}$-forces
become strong at $\Lambda_{NP}= 10 TeV$, and therefore a related NP should
appear by the time we reach this energy scale.

After establishing our expectations on how large these NP couplings
could be, we turn to the process $e^{-}e^{+}\to
HZ_{(f\bar f)}$. Its total cross section as a function of the
$e^{-}e^{+}$
energy is shown in Fig.1 assuming $m_H=80GeV$. This value has been chosen
optimistically, hoping that the Higgs will be produced at LEP2
through $e^-e^+ \to ZH$. In this
figure, we present results for various values of the $d$ and $d_B$ couplings,
indicating also the corresponding NP scale they correspond to. Identical
results are of course also obtained for the same values of the
$CP$-violating couplings $\overline{d}$ and $\overline{d}_B$ respectively.

The $HZ$ angular distribution is mainly sensitive to $f_{\Phi 2}$ through
the renormalization factor $Z_H$ and to the combination
$d_{ZZ}=dc_W^2+d_Bs_W^2$. Figs.2a,b,c show this sensitivity for the three
nominal LEP2 energies, 175, 192 and 205 $GeV$. Assuming an integrated
luminosity of 500, 300 and 300 $pb^{-1}$ respectively,
one expects about two hundred
raw events; this number being then reduced by the $H$ and $Z$ detection.
Because of this, we cannot expect to measure this cross section at LEP2 with
an accuracy better than about 10\%. Assuming that only one NP operator acts
each time, we then deduce an observability limit $|f_{\phi 2}|\simeq 0.01$
corresponding to $\Lambda _{NP}\simeq 6-7TeV$, or an
observability limit $|d|\simeq 0.015$ ( $|d_B|\simeq 0.05$)
corresponding to $\Lambda _{NP}\simeq
6-7TeV$ ( $\Lambda _{NP}\simeq 5TeV$ ) for $m_H\simeq 80GeV$. This is not so
bad for a first exploration of the Higgs sector! However, with such a number
of events it is not conceivable to make a meaningful analysis of the $Z$
decay distribution in order to disentangle the various NP operators.

At higher energies, with the designed luminosities of NLC, the number of
events is increasing as well as the sensitivity to anomalous couplings. For
example at 1 $TeV$, with 80 $fb^{-1}$ per year, the SM rate
gives one thousand
events per year. One can then expect a measurement of the cross section with
a 3\% accuracy. The implied sensitivities to the various couplings
become $|f_{\phi 2}|\simeq 0.004$, $|d|\simeq 0.005$ and $|d_B|\simeq 0.015$
corresponding to NP scales of 10, 11 and 9 $TeV$ respectively;
compare Fig.1 and Fig.3. Moreover, with such a number of events, azimuthal
asymmetries could be used in order to ~disentangle the various $d$-type
couplings.

As noted already, there are four kinds of such asymmetries. We first discuss
$A_{14}$ and $A_{12}$ which are shown in Figs.4,5. These asymmetries depend
on the flavour of the fermion $f$ to which $Z$ decays and are mostly
interesting for the $Z\to b\bar b$ case. From Figs.4,5 it looks possible
that the sensitivities to $d$ and $d_B$ could reach the percent level. Note
also that $A_{14}$, $A_{12}$ depend mainly on the combination $d_{\gamma
Z}=s_Wc_W(d-d_B)$ and on its $CP$-violating partner $\overline{d}_{\gamma Z}$%
.Thus, the illustrations made for $d$ ($\overline{d}$)  apply also to $-d_B$%
, ($-\overline{d}_B$). The asymmetry $A_{13}$ shown in Fig.6 does not depend
on the type of the fermion $f$ but it is sensitive to the $CP$-violating
combination $\overline{d}_{ZZ}$; (compare (8)). Thus,~sensitivities to this
coupling at the percent level should be possible using this asymmetry. Note
that in this case the sensitivities to $d_B$ and $d$ are related by $%
d_B\simeq dc_W^2/s_W^2$. Finally the asymmetry $A_8$ shown in Fig.7, is also
mainly sensitive to $d_{ZZ}$ and independent of the flavour of the $f$
fermion. Consequently, the ~differential cross section for
$e^{-}e^{+}\to ZH$, together
with the above asymmetries, provide considerable information for
disentangling the aforementioned five relevant anomalous
couplings at the percent level.
It follows that NP scales of the order of a few $TeV$ could be
searched for this way.

Further information on the above four $d$-type couplings, can be obtained
from the ~differential cross section for the process
$e^{-}e^{+}\to \gamma H$.
At SM this cross section is unobservably small, but it is very sensitive
to the four NP interactions. Thus, the differential cross section could
become observable if non negligible anomalous interactions occur,
which according to (30) imply an angular distribution of the form $1+\cos
^2\vartheta $ \cite{Maettig}.  If $m_H \lsim 90GeV$, this process is
 allowed already at LEP1, leading to a rather low energy photon
together with a Higgs particle decaying mainly to $b\bar b$.
The background to this process comes mainly  from final
state radiation in the $Z\to b\bar b$ process. A discussion of this
problem has been given in \cite{HLEP, Maettig}.
On the basis of this we conclude that for \eg\@
$m_H \sim 80GeV$, we would need about a hundred events in order
to obtain a visible signal over a background of about $10^{+7}$
$Z$. This then means that $\sigma(e^-e^+ \to \gamma H) \gsim
0.3pb$ is required, which is at least a
factor 100 above the SM rate and requires anomalous Higgs couplings
satisfying $|d|\gsim 0.1$ or $|d_B| \gsim 0.1$. So these (rough)
constraints leave room for the analysis of
the $e^-e^+\to HZ$ process
at LEP2 that we proposed above.\par

One can then look for the $e^{-}e^{+}\to \gamma H$ process at
LEP2 where the Higgs is associated to very energetic
photons, for which there should be no such background.
As a first illustration we plot in Fig.8a the
photon angular distribution in terms of $|\cos \vartheta |$,
for $m_H = 80GeV$ and $e^-e^+$ c.m. energy $192GeV$.
There are various remarks to be made now. Thus, for SM we
find $\sigma(e^-e^+ \to \gamma H) \simeq
0.16fb$, which is consistent with the result of\footnote{It is
 amusing to state that the tools
for the numerical calculations used here are rather primitive compared to
\cite{tools} employed in \cite{HgAA}.} \cite{HgAA}. As we see
from this figure, a much higher cross section will appear if an
anomalous interaction is present. Because of this and demanding
a few events,
one sees that a sensitivity to $|d|\simeq 0.05$ or $|d_B|\simeq 0.025$ is
possible for $m_H\sim 80GeV$, which means testing  NP scales
of 3 and 7 $TeV$ respectively.  These later sensitivities come
from ($d$, $d_B$) combinations that are
different from those appearing in $HZ$ production,
which provides a welcome
help for disentangling these two operators. Since
$m_H=80GeV$ is far below the LEP2 c.m.\@ energy, we observe
that the cross
sections presented in Fig.8a remain true also for the other LEP2
nominal energies, ~namely $175$ and $205GeV$. Finally we should
remark that the difference between the
predictions for the various cross sections corresponding to
opposite signs of any anomalous
coupling, gives a measure of the magnitude of the ~interference
contribution between the standard and anomalous terms. It can
therefore be seen that \eg\@ in Fig.8a, this difference is much
smaller than the difference  in Figs.2a,b,c.\par

The very interesting thing concerning $e^{-}e^{+}\to
\gamma H$ is that it allows for a similar sensitivity on $d$ or
$d_B$, even in the case that the Higgs is considerably
heavier. This can be seen from Fig.8b where the angular
distribution for \eg\@ $m_H=120GeV$ and a total
$e^-e^+$ energy of $192GeV$, is presented. It can also be seen
there, that  $\sigma(e^-e^+
\to \gamma H) \sim 13fb$ for $|d| \sim 0.1$, while the SM expectation
is about $0.084fb$ \cite{HgAA}. \par

For a $1TeV$ NLC, Fig.9a,b show a sensitivity to $|d|\simeq
0.005$ and $|d_B|\simeq 0.0025$ corresponding to NP scales of 11 and 22
$TeV$ respectively. In these figures we have used for
illustration $m_H=80GeV$, but of course similar results would
have been expected for any $m_H \ll 1TeV $.
Note also that  at such an energy, the SM differential cross section
is somewhat enhanced in the forward and backward regions, while
the total cross section reaches the $0.014fb$ level.

\section{Conclusions}

We have studied the sensitivity to NP of the processes $e^{-}e^{+}\to ZH$
and $\gamma H $ that are observable at LEP2 and higher energy
$e^{-}e^{+}$
colliders. Our analysis is concentrated on the residual NP effects described
by the $dim=6$ $SU(2)\times U(1)$ gauge invariant operators ${\cal O}_{UB}$,
${\cal O}_{UW}$, $\overline{{\cal O}}_{UB}$, $\overline{{\cal
O}}_{UW}$, ${\cal O}_{\Phi 2}$, which only affect the Higgs
couplings to themselves and to
gauge bosons. We have given explicit analytic expressions for the helicity
amplitudes and the angular distributions for $HZ$ and $H\gamma $
as well as $Z$ the spin density matrix elements and
azimuthal asymmetries in $Z\to f\bar
f$ decays.

We  first performed an analysis for the LEP2 conditions, using $M_H=80GeV$
as an illustration. From the study of $d\sigma /d\cos \vartheta $ in $%
e^{-}e^{+}\to ZH$, a sensitivity to ${\cal O}_{\Phi 2}$ is
expected up to a NP
scale of 6-7 $TeV$, and to ${\cal O}_{UB}$ and ${\cal O}_{UW}$ up to scales
of 5 and 7 $TeV$ respectively. For the process $e^{-}e^{+}\to
\gamma H$ the SM rate is
unobservable, but the anomalous interactions we consider could
strongly enhance it. At LEP1 the background is too strong to allow for the
observability of this process if $m_H \sim 80 GeV$ and the NP scales
are at the aforementioned level. At LEP2, where the background is reduced,
the observation of a few events would imply the presence
of NP due to ${\cal O}_{UB}$ or ${\cal O}_{UW}$ effects at 7 or 3 $TeV$
respectively. Furthermore, a comparison between the two processes
$e^-e^+ \to ZH$ and $e^-e^+ \to \gamma H$ could be used for
disentangling the effects of the operators ${\cal O}_{UB}$
and ${\cal O}_{UW}$ that are involved in different
combinations.\par

The $e^-e^+ \to \gamma H$ process is also
particularly interesting at LEP2 energies, as it
may allow for the production and study of a Higgs particle with
$m_H \gsim 100GeV$, which would not be  possible through
$e^-e^+ \to ZH$. Thus, for \eg\@ $m_H=120GeV$ and a total
$e^-e^+$ energy of $192GeV$, we should be able to observe NP effects
at scales similar to those observable in the $m_H \sim 80GeV$ case.\par

At NLC in the $1 TeV$ range, the sensitivity increases due to the
$s$-dependence and the larger luminosity. Studying therefore
$e^{-}e^{+}\to ZH$, a
complete disentangling of the five operators can be
envisaged by using in
addition the four azimuthal asymmetries associated to the $\cos
\phi _f$, $\sin \phi _f$, $\sin 2\phi _f$ and $\cos 2\phi _f$
distributions. This
disentangling should be achievable down to the percent level for the NP
couplings, implying sensitivities to NP scales of the
order of 10, 8 and 12 $TeV$ for ${\cal O}_{\Phi 2}$, ${\cal
O}_{UB}$ and ${\cal O}_{UW}$ respectively. The $e^{-}e^{+}\to \gamma
H$ process could feel ${\cal O}_{UB}$ and ${\cal O}_{UW}$ up to scales of 20
and 12 $TeV$. In this respect the disentangling may be further helped by
comparing with the analyses of other bosonic processes like {\it e.g.\/}
boson pair production \cite{ggVV1, ggVV2} and single Higgs boson production
in $\gamma \gamma $ collisions \cite{higpro1, higpro2}.

In conclusion, if $m_H$ is small enough, we can expect a remarkable first
exploration of the Higgs sector at LEP2, while at NLC a detail study of the
NP properties should be made possible for a wider range of $m_H$ and $%
\Lambda _{NP}$ scales.

{\it ~Acknowledgements}: We would like to thank Jorge Rom\~{a}o and Wayne
Repko for communications concerning Refs. \cite{Barro, HgAA}.

\newpage

\renewcommand{\theequation}{A.\arabic{equation}} \setcounter{equation}{0} %
\setcounter{section}{0}

{\large {\bf Appendix A : Z spin density matrix elements in $e^{-}e^{+}\to ZH
$}} From the helicity amplitudes of Section 2 one computes the
(un-normalized) Z density matrix elements
\begin{equation}
\rho _{\tau \tau ^{\prime }}=\rho _{\tau ^{\prime }\tau }^{*}=\sum_{\lambda
=\pm 1}F_{\lambda \tau }F_{\lambda \tau ^{\prime }}^{*}\ \ .\
\end{equation}
It is convenient to define the following left-handed and right-handed
combinations of couplings:
\begin{eqnarray}
D_{+} &=&s_Wc_W\Big [ d(1-\chi )-d_B(1+\chi \frac{s_W^2}{c_W^2})\Big ] \ \
,\  \\
D_{-} &=&s_Wc_W\Bigg [ d\left( 1+\chi (\frac 1{2s_W^2}-1)\right) -d_B\left(
1+\chi (\frac 1{2c_W^2}-1)\right) \Bigg ] \ \ ,\
\end{eqnarray}
and similarly for the $CP$-violating couplings. As before, $\chi =s/(s-M_Z^2)
$. Using these ~definitions and denoting by $Z_H$ the $H$ field wave
function ~renormalization induced by ${\cal O}_{\Phi 2}$, the Z density
matrix elements become
\begin{eqnarray}
\rho _{++}(\vartheta ) &=&Z_H\,{\frac{e^2g^2(1+\cos ^2\vartheta )}{2sM_W^2}}%
\Bigg \{ M_Z^4\chi ^2{\frac{(1-4s_W^2+8s_W^4)}{4s_W^2c_W^2}}  \nonumber \\
&&+{\frac{2M_Z^2\chi k_0\sqrt{s}}{s_Wc_W}}~[(1-2s_W^2)D_{-}-2s_W^2D_{+}]
\nonumber \\
&&+4k_0^2s(D_{+}^2+D_{-}^2)+4k^2s(\overline{D}_{+}^2+\overline{D}_{-}^2)%
\Bigg \}  \nonumber \\
&&-Z_H{\frac{e^2g^2\cos \vartheta }{sM_W^2}}\Bigg \{ M_Z^4\chi ^2{\frac{%
(1-4s_W^2)}{4s_W^2c_W^2}}+{\frac{2M_Z^2\chi k_0\sqrt{s}}{s_Wc_W}}%
[(1-2s_W^2)D_{-}+2s_W^2D_{+}]  \nonumber \\
&&+4k_0^2s(D_{-}^2-D_{+}^2)+4k^2s(\overline{D}_{-}^2-\overline{D}_{+}^2)%
\Bigg \} \ \ ,
\end{eqnarray}
\begin{equation}
\rho _{--}(\vartheta )~=~\rho _{++}(\pi -\vartheta )\ \ ,
\end{equation}
\begin{eqnarray}
\rho _{00}(\vartheta ) &=&Z_H\,{\frac{e^2g^2\sin ^2\vartheta }{sc_W^2}}~%
\Bigg \{ {\frac{k_0^2\chi ^2}{4c_W^2s_W^2}}~(1-4s_W^2+8s_W^4)  \nonumber \\
&&-{\frac{2\sqrt{s}k_0\chi }{c_Ws_W}}%
{}~[2s_W^2D_{+}-(1-2s_W^2)D_{-}]+4s[D_{+}^2+D_{-}^2]\Bigg \} \ \ \ ,\
\end{eqnarray}
\begin{eqnarray}
\rho _{0+}(\vartheta ) &=&Z_H\,{\frac{e^2g^2\sin \vartheta }{\sqrt{2}M_W^2s}}%
{}~\Bigg \{ {\frac{M_Z^4\chi ^2k_0}{4s_W^2c_WM_W}}[1-4s_W^2-(1-4s_W^2+8s_W^4)%
\cos \vartheta ]  \nonumber \\
&&+{\frac{M_Z^3\sqrt{s}\chi }{s_Wc_W}}\Big(1+{\frac{k_0^2}{M_Z^2}}\Big) %
{}~[(1-2s_W^2)(1-\cos \vartheta )D_{-}+2s_W^2(1+\cos \vartheta )D_{+}]
\nonumber \\
&&+4sM_Zk_0[(1-\cos \vartheta )D_{-}^2-(1+\cos \vartheta )D_{+}^2]  \nonumber
\\
&&+4ikM_Zs[D_{-}\overline{D}_{-}(1-\cos \vartheta )-D_{+}\overline{D}%
_{+}(1+\cos \vartheta )]  \nonumber \\
&&+i~{\frac{kk_0\sqrt{s}M_Z\chi }{c_Ws_W}}~[(1-\cos \vartheta )(1-2s_W^2)%
\overline{D}_{-}+2s_W^2(1+\cos \vartheta )\overline{D}_{+}]\Bigg \} \ \ ,\
\end{eqnarray}
\begin{equation}
\rho _{0-}(\vartheta )=\rho _{0+}^{*}(\pi -\vartheta )\ \ ,
\end{equation}
\begin{eqnarray}
\rho _{+-}(\vartheta ) &=&Z_H\,{\frac{e^2g^2\sin ^2\vartheta }{2M_W^2s}}~%
\Bigg \{ M_Z^4\chi ^2~{\frac{(1-4s_W^2+8s_W^4)}{4s_W^2c_W^2}}  \nonumber \\
&&+{\frac{2M_Z^2\chi k_0\sqrt{s}}{s_Wc_W}}~[(1-2s_W^2)D_{-}-2s_W^2D_{+}]
\nonumber \\
&&+4k_0^2s(D_{+}^2+D_{-}^2)-4k^2s(\overline{D}_{+}^2+\overline{D}_{-}^2)
\nonumber \\
&&-{\frac{2ik\sqrt{s}M_Z^2\chi }{s_Wc_W}}~[(1-2s_W^2)\overline{D}_{-}-2s_W^2%
\overline{D}_{+}]  \nonumber \\
&&-8ikk_0s(D_{+}\overline{D}_{+}+D_{-}\overline{D}_{-})\Bigg \}\ \ .\
\end{eqnarray}

\newpage

\renewcommand{\theequation}{B.\arabic{equation}} \setcounter{equation}{0} %
\setcounter{section}{0}

{\large {\bf Appendix B : Z decay distributions and projectors}}

\vspace{0.5cm}

Assuming that the $Z\to f \bar f$ amplitude is standard, the angular
distribution of the $f\bar f$ ~system in the $Z$ rest frame is determined
according to (17) by $Z$ density matrix $\rho_{\tau \tau^{\prime}}$ in the
helicity basis and the decay elements $Q_{\tau\tau^{\prime}}$. Using the
notation introduced in Sect.3 and the definitions immediately after (9) we
have
\begin{equation}
Q_{\pm\pm} = {\frac{3}{16\pi}}\left (1+\cos^2\theta_f \mp {\frac{4A_f V_f}{%
A^2_f+V^2_f}}\, \cos\theta_f \right ) \ \ ,
\end{equation}
\begin{equation}
Q_{00} = {\frac{3}{8\pi}}\sin^2\theta_f \ \ ,
\end{equation}
\begin{equation}
Q_{\pm\mp} = {\frac{3}{16\pi}}\sin^2\theta_f e^{\pm2i\phi_f} \ \ ,
\end{equation}
\begin{equation}
Q_{\pm 0} = Q^*_{0 \pm} = {\frac{3\sqrt{2}}{16\pi}}\sin\theta_f e^{\pm
i\phi_f} \left [\pm \cos\theta_f-{\frac{2A_f V_f}{A^2_f +V^2_f}} \right ] \
\ .
\end{equation}
The associated $\Lambda^Z_{\tau\tau^{\prime}}$ projectors defined by
\begin{equation}
\int Q_{\tau\tau^{\prime}}(\theta_f,\phi_f)\Lambda^Z_
{\mu\mu^{\prime}}(\theta_f,\phi_f)d\Omega_f =
\delta_{\tau\mu}\delta_{\tau^{\prime}\mu^{\prime}} \ \ ,
\end{equation}
are
\begin{equation}
\Lambda^Z_{00} = 2-5\cos^2\theta_f \ \ ,
\end{equation}
\begin{equation}
\Lambda^Z_{\pm\pm} = {\frac{1}{2}}\left [5\cos^2\theta_f \mp{\frac{%
V^2_f+A^2_f}{A_f V_f}}\cos\theta_f -1 \right ] \ \ ,
\end{equation}
\begin{equation}
\Lambda^Z_{\pm\mp} = 2e^{\mp2i\phi_f} \ \ ,
\end{equation}
\begin{equation}
\Lambda^Z_{\pm 0} = \Lambda^{Z*}_{0\pm} = -{\frac{2\sqrt2}{3\pi}}~\left [{%
\frac{V^2_f+A^2_f}{V_f A_f}} \mp 8\cos\theta_f\right ] e^{\mp i\phi_f} \ \ .
\end{equation}

The $Z$ density matrix elements can be extracted from a given angular
distribution by integrating over the final fermion solid angle using these
projectors according to
\begin{equation}
\rho_{\tau\tau^{\prime}}(\vartheta) = {\frac{64\pi s\sqrt{s}}{kB(Z\to f\bar
f)}}\int d\Omega_f {\frac{d \sigma(e^+e^-\to Hf\bar f) }{d \cos\vartheta d
\Omega_f}}\Lambda^Z_{\tau\tau^{\prime}}\ \ .
\end{equation}

\newpage

\centerline {\ {\bf Figure Captions }}

Fig.1 Cross section for $e^+e^-\to HZ$ versus $e^+e^-$ total energy for $%
m_H=80 GeV$. SM and NP contributions due to $ {\cal O }_{UB}$, $ {\cal O }%
_{UW}$, and $ {\cal O }_{\Phi2}$.\\

Fig.2 Angular distribution of $e^+e^-\to HZ$ versus $|cos\vartheta|$ at
three LEP2 energies (a) 175 GeV, (b) 192 GeV, (c) 205 GeV, for $m_H=80 GeV$,
SM and NP contributions due to $ {\cal O }_{UB}$, $ {\cal O }_{UW}$, and $
{\cal O }_{\Phi2}$. The number of events obtained with an integrated
luminosity of 500, 300, 300 $pb^{-1}$ respectively, is also indicated.\\

Fig.3 Angular distribution of $e^+e^-\to HZ$ versus $|cos\vartheta|$ at NLC
(1 TeV), for $m_H=80 GeV$, SM and NP contributions due to $ {\cal O }_{UB}$,
$ {\cal O }_{UW}$, and $ {\cal O }_{\Phi2}$. For $f_{\Phi2}=0.01$ the curve
coincides with the one for $d=-0.01$. The number of events obtained with an
integrated luminosity of 80 $fb^{-1}$ is also indicated.\\

Fig.4 Azimuthal asymmetry $A_{14}$ versus total $e^+e^-$ energy for $m_H=80
GeV$, SM and NP contributions due to $ {\cal O }_{UW}$, (a) $Z\to b\bar b$,
(b)$Z\to \mu^+\mu^-$.\\

Fig.5 Azimuthal asymmetry $A_{12}$ versus total $e^+e^-$ energy for $m_H=80
GeV$, SM and NP contributions due to $\overline{{\cal O }}_{UW}$, (a) $Z\to
b\bar b$, (b)$Z\to \mu^+\mu^-$.\\

Fig.6 Azimuthal asymmetry $A_{13}$ versus total $e^+e^-$ energy for $m_H=80
GeV$, SM and NP contributions due to $\overline{{\cal O }}_{UW}$.\\

Fig.7 Azimuthal asymmetry $A_{8}$ versus total $e^+e^-$ energy for $m_H=80
GeV$, SM and NP contributions due to $ {\cal O }_{UW}$.\\

Fig.8 Angular distribution of $e^+e^-\to H\gamma$ versus $|cos\vartheta|$ at
the LEP2 energy $192 GeV$, for $m_H=80 GeV$ (a), or $m_H=120GeV$
(b). The SM and NP contributions are due to $ {\cal O }_{UW}$.
For $ {\cal O }_{UB}$ similar results are obtained
provided $d_B \sim d/2$. The number of events
obtained with an integrated luminosity of  $300 pb^{-1}$
is also indicated.\\

Fig.9 Angular distribution of $e^+e^-\to H\gamma$ versus $|cos\vartheta|$ at
NLC (1 TeV), for $m_H=80 GeV$, (a) SM and NP contributions due to $ {\cal O }%
_{UW}$, (b) SM and NP contributions due to $ {\cal O }_{UB}$. The number of
events obtained with an integrated luminosity of 80 $fb^{-1}$ is also
indicated.\\

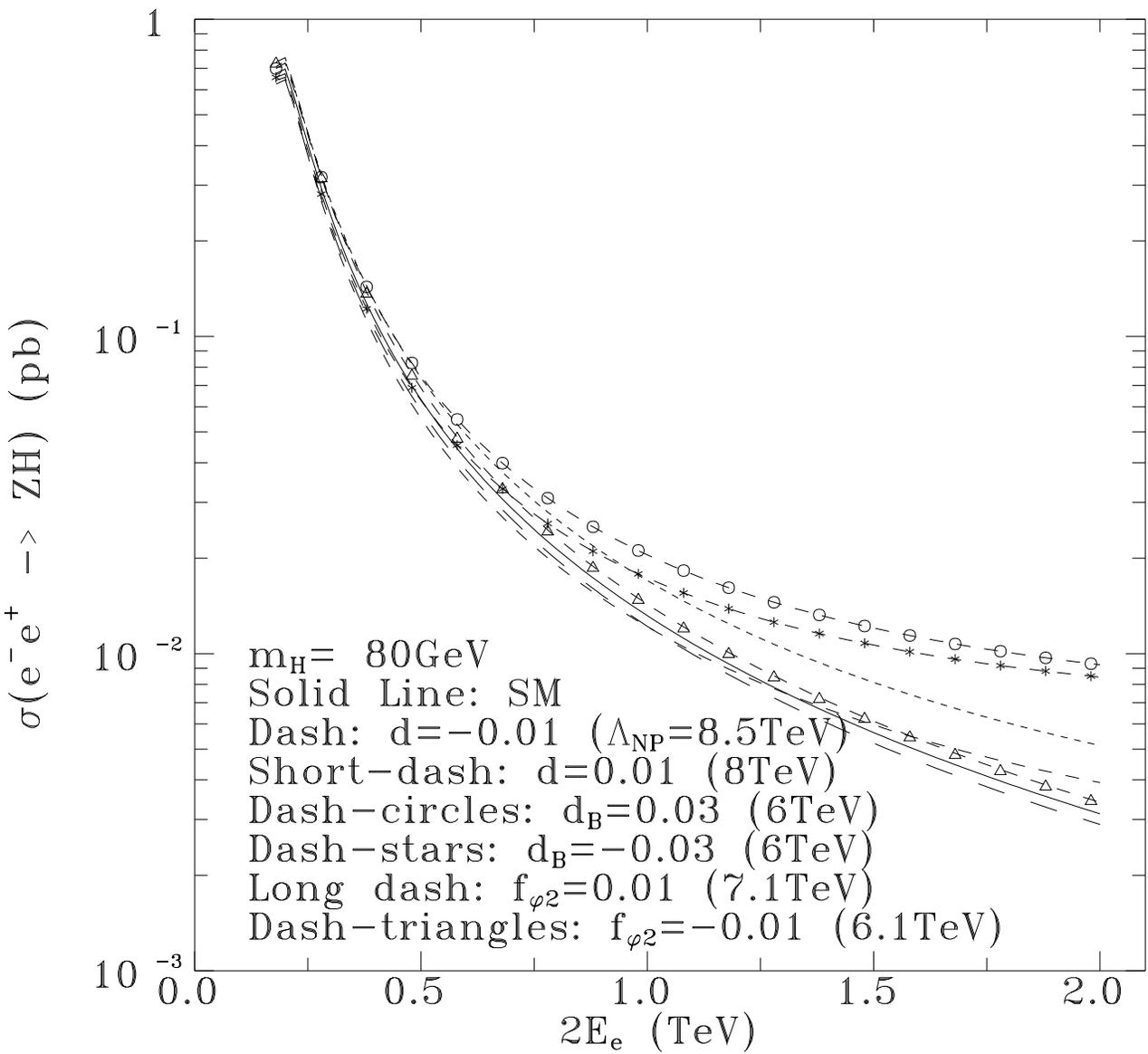
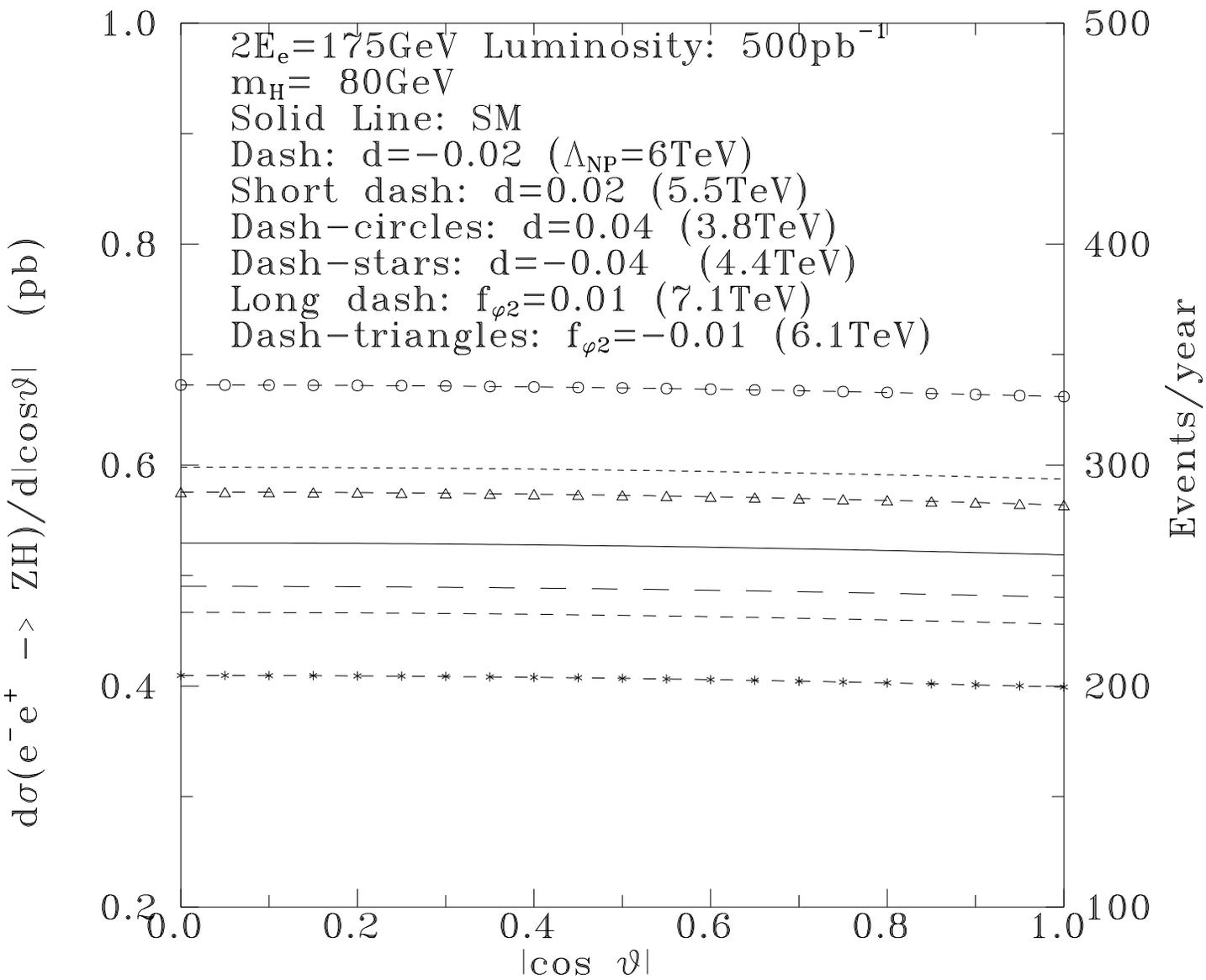
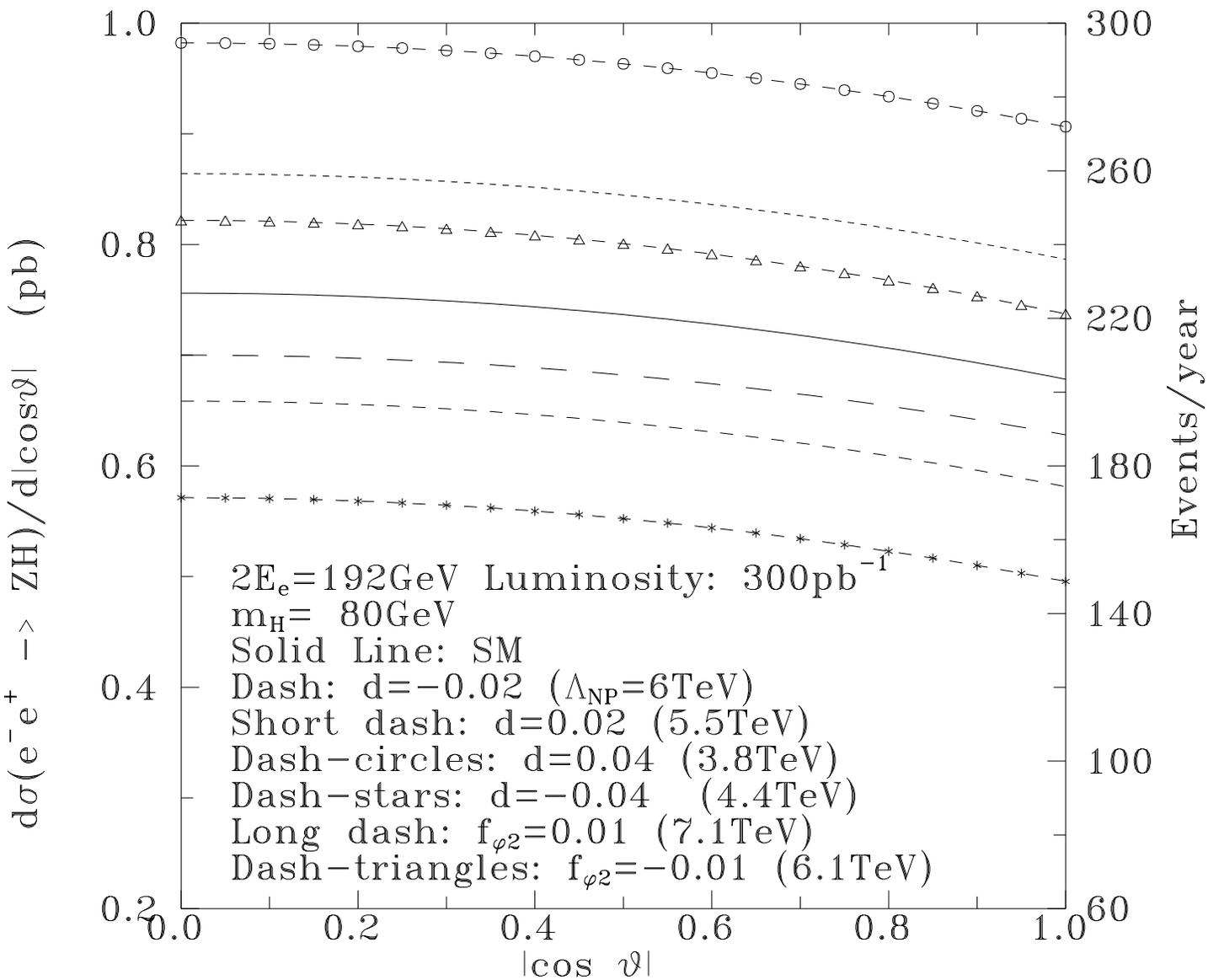
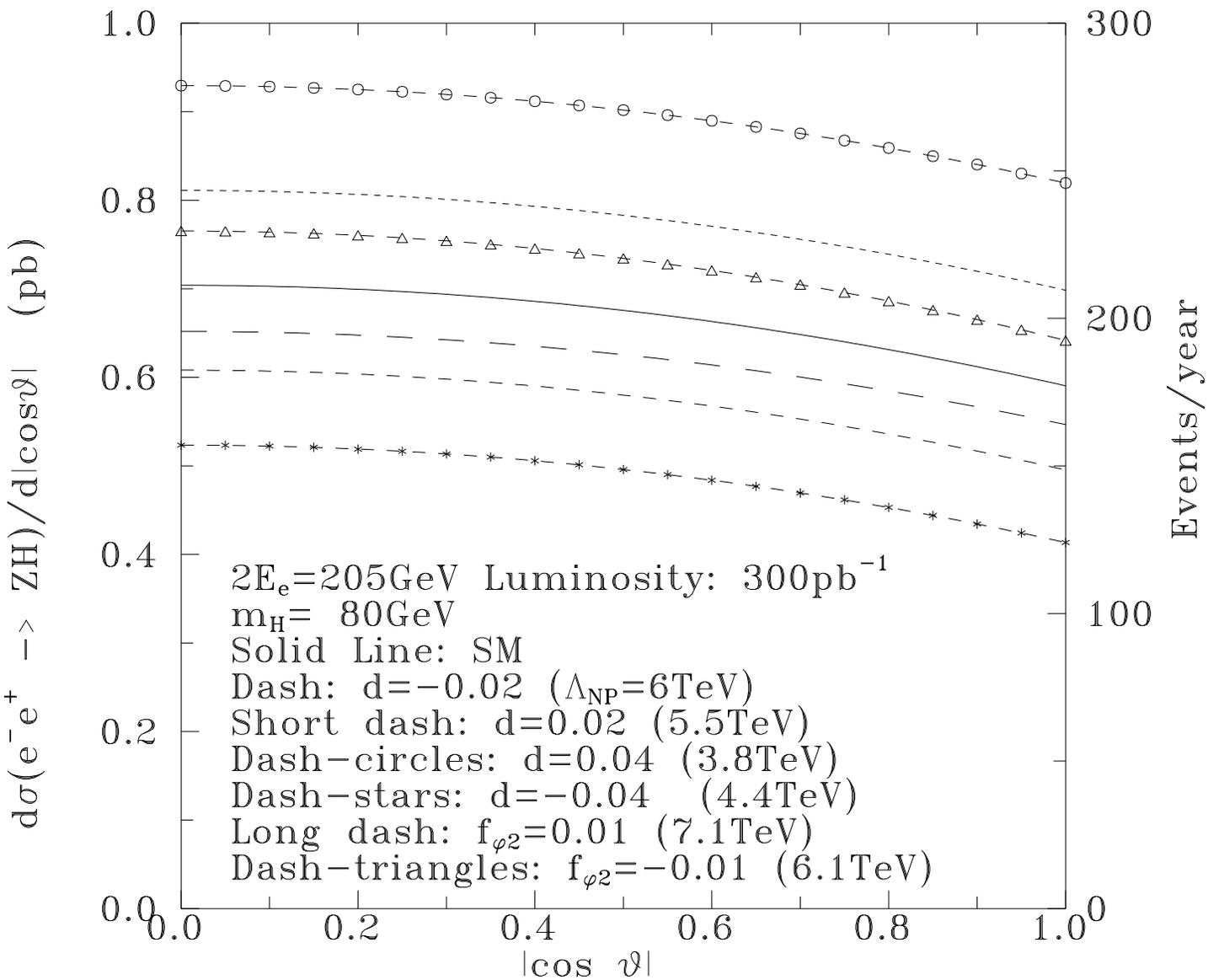
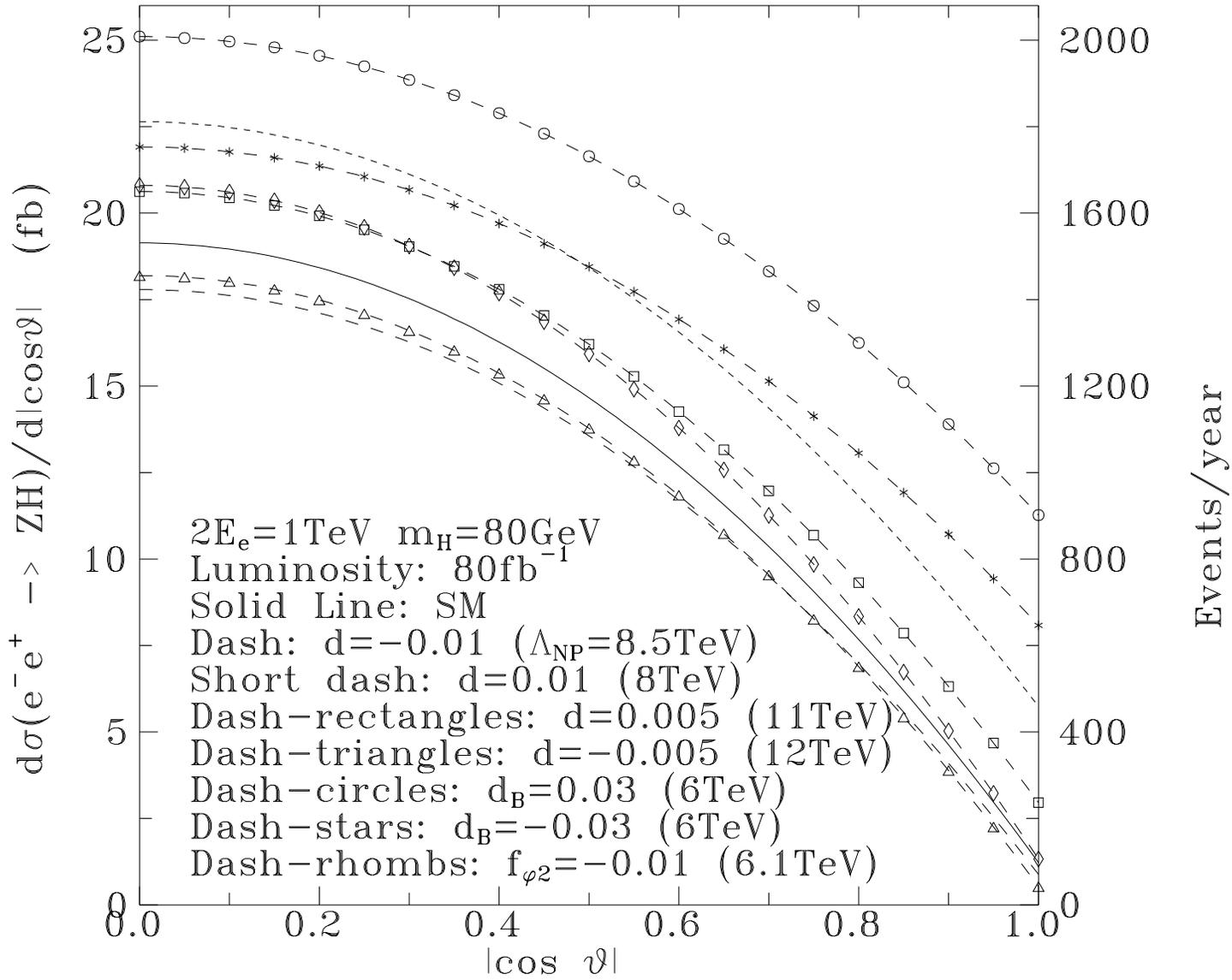
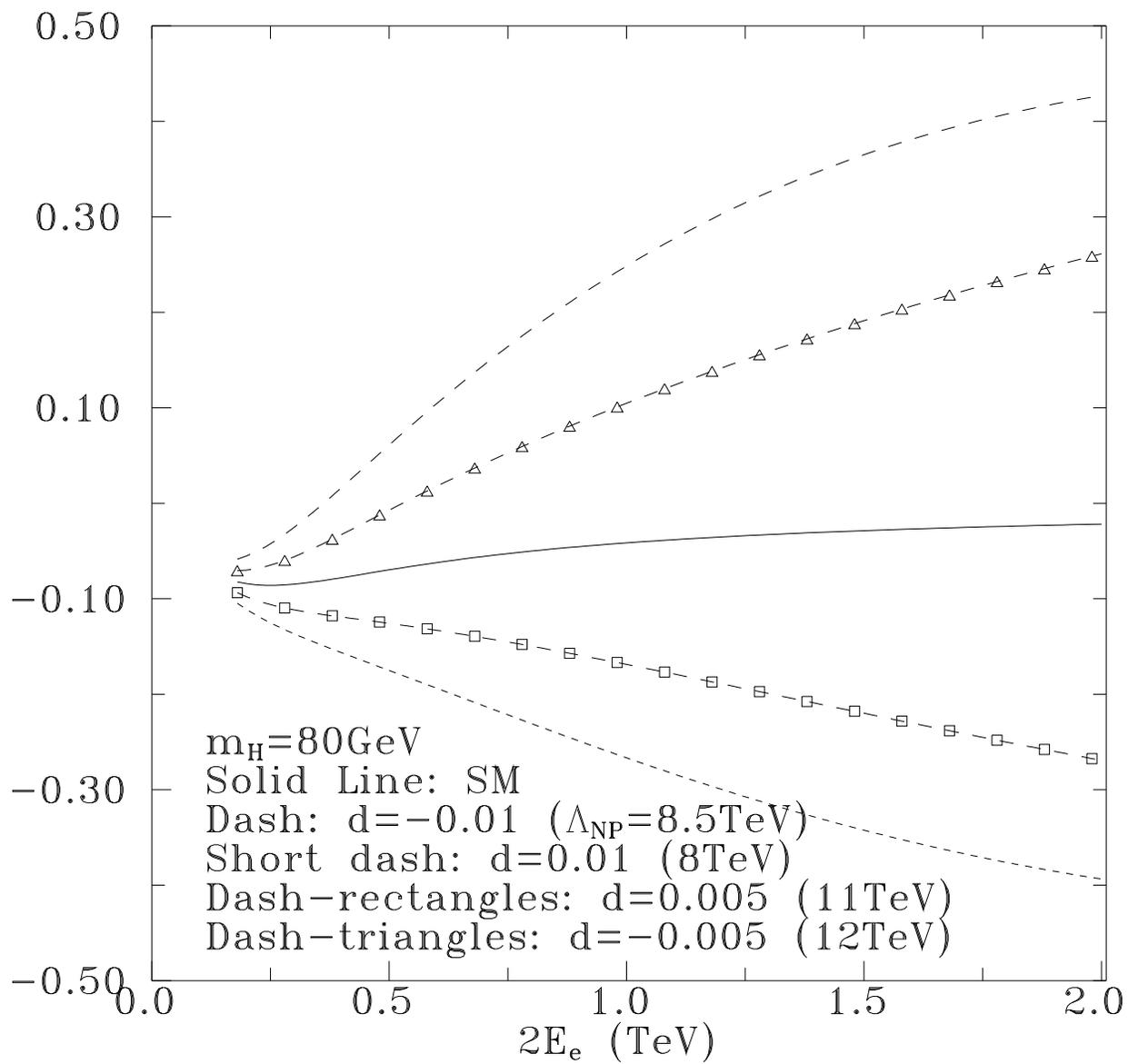
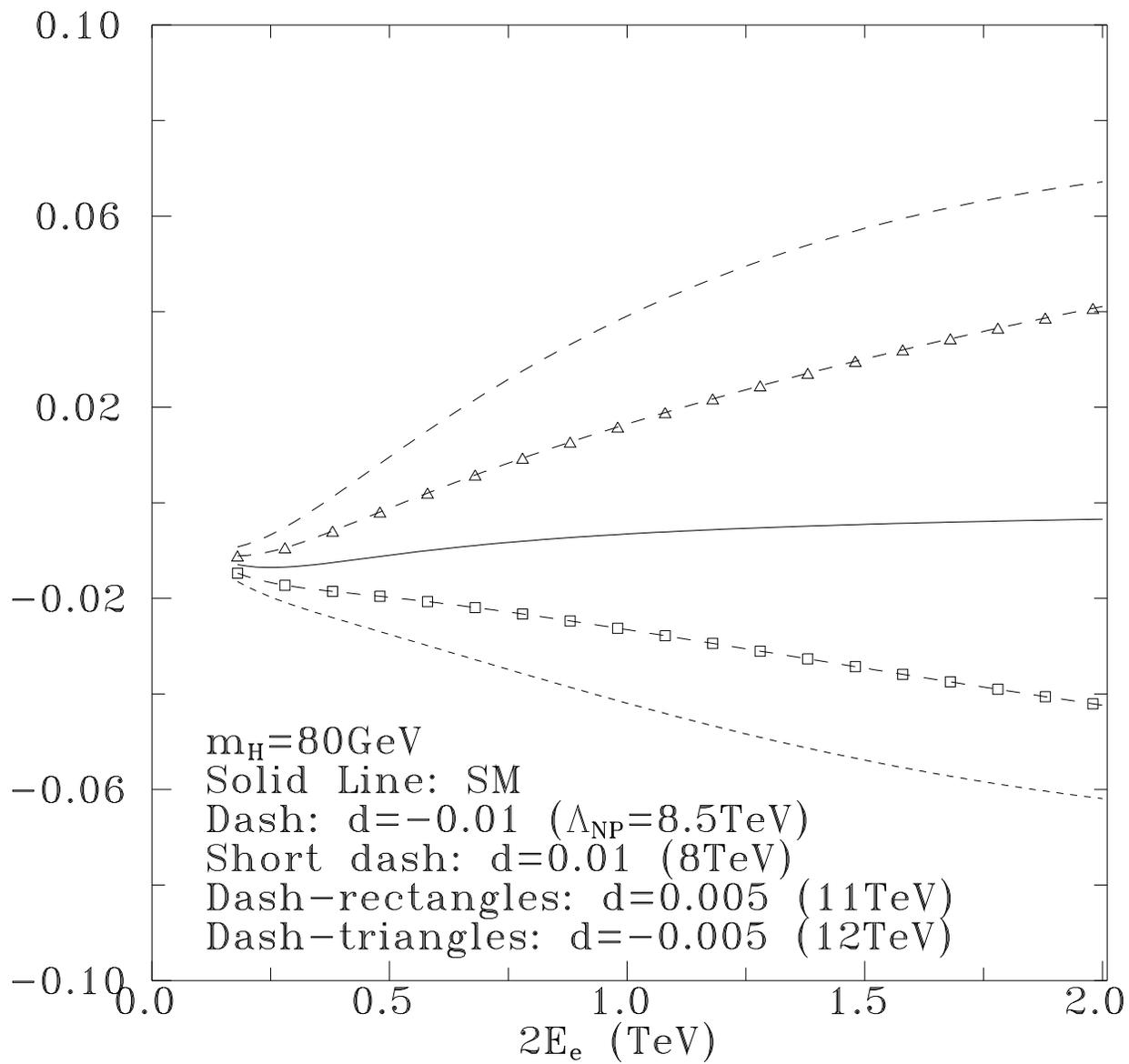
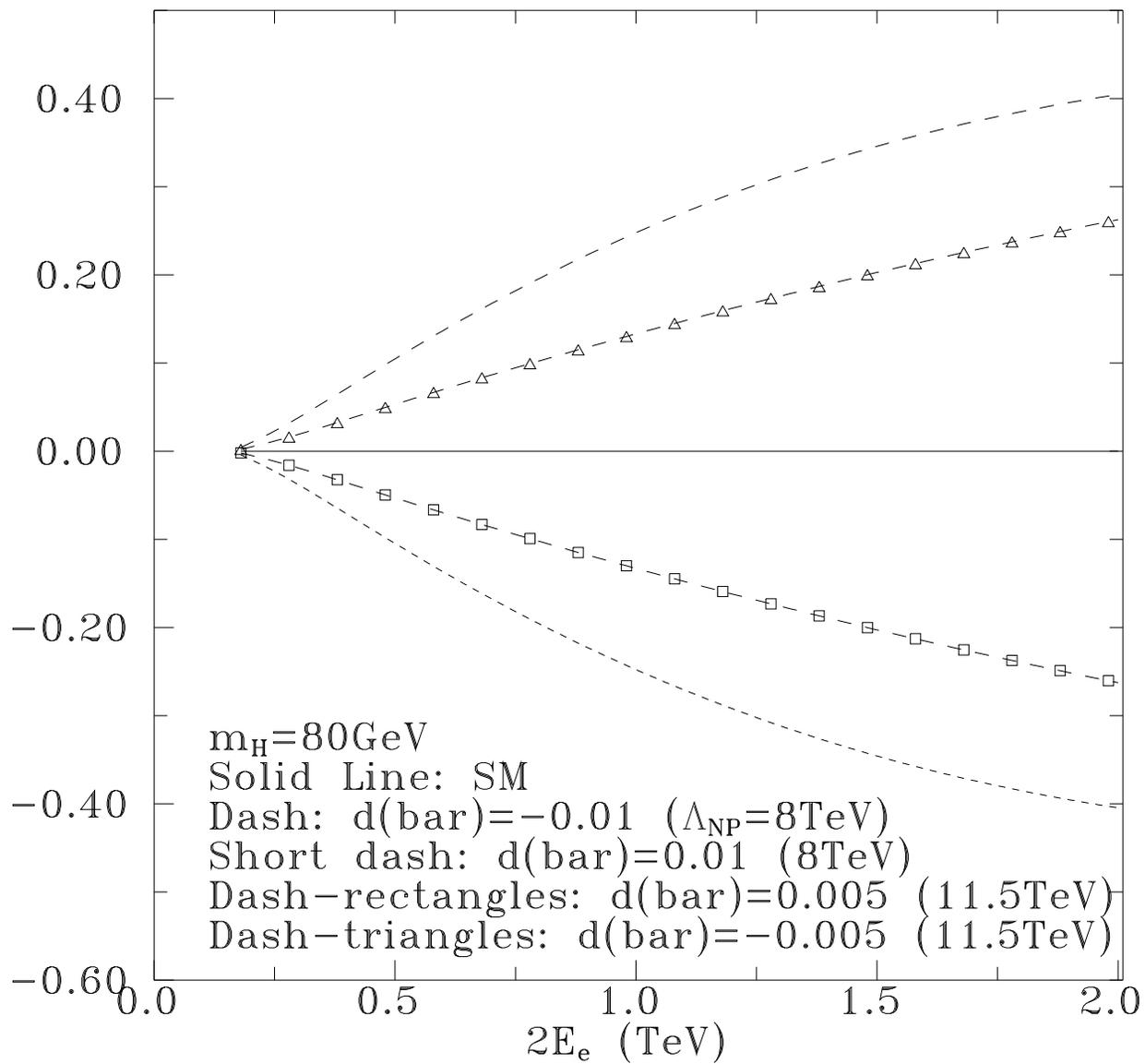
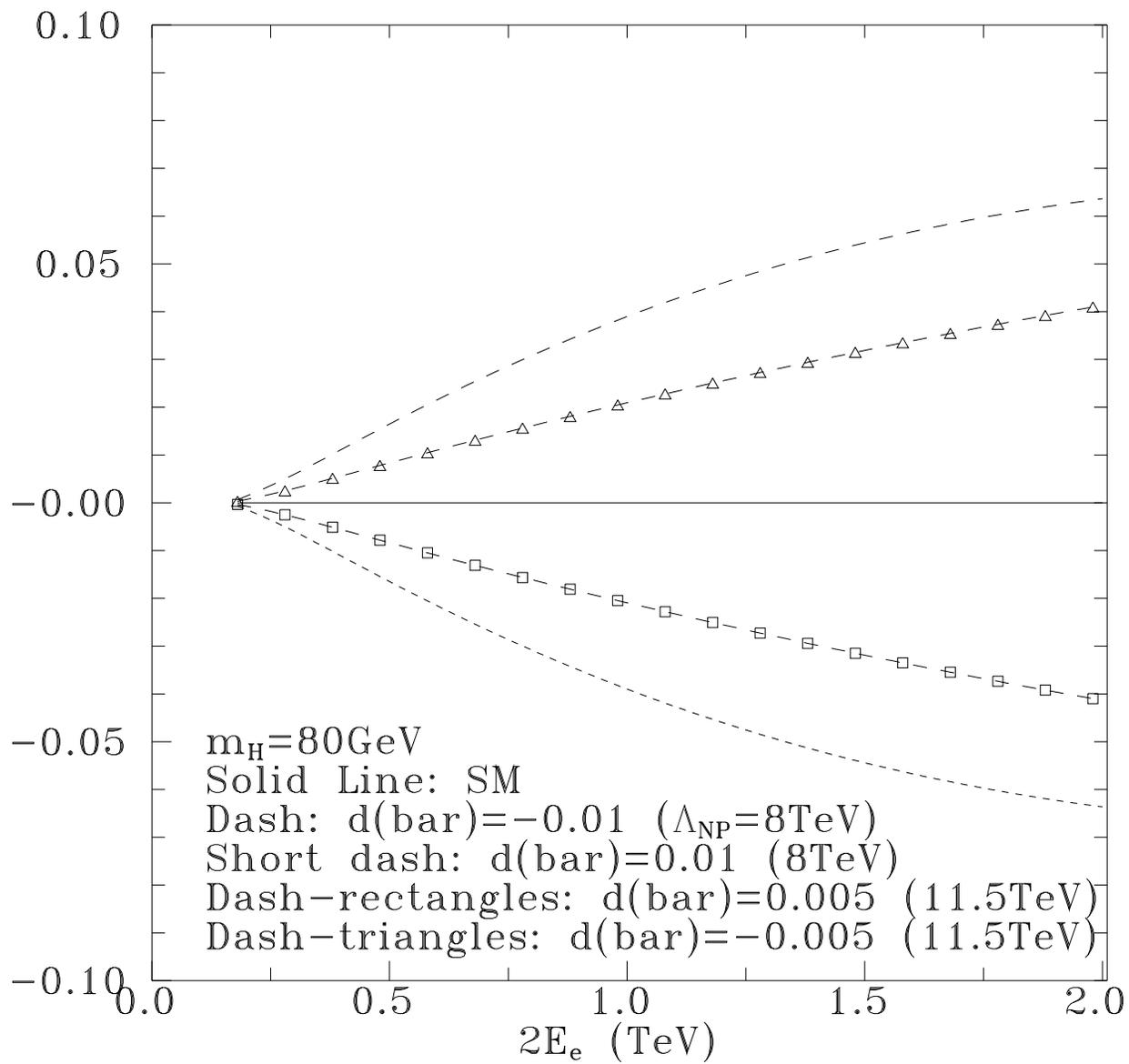
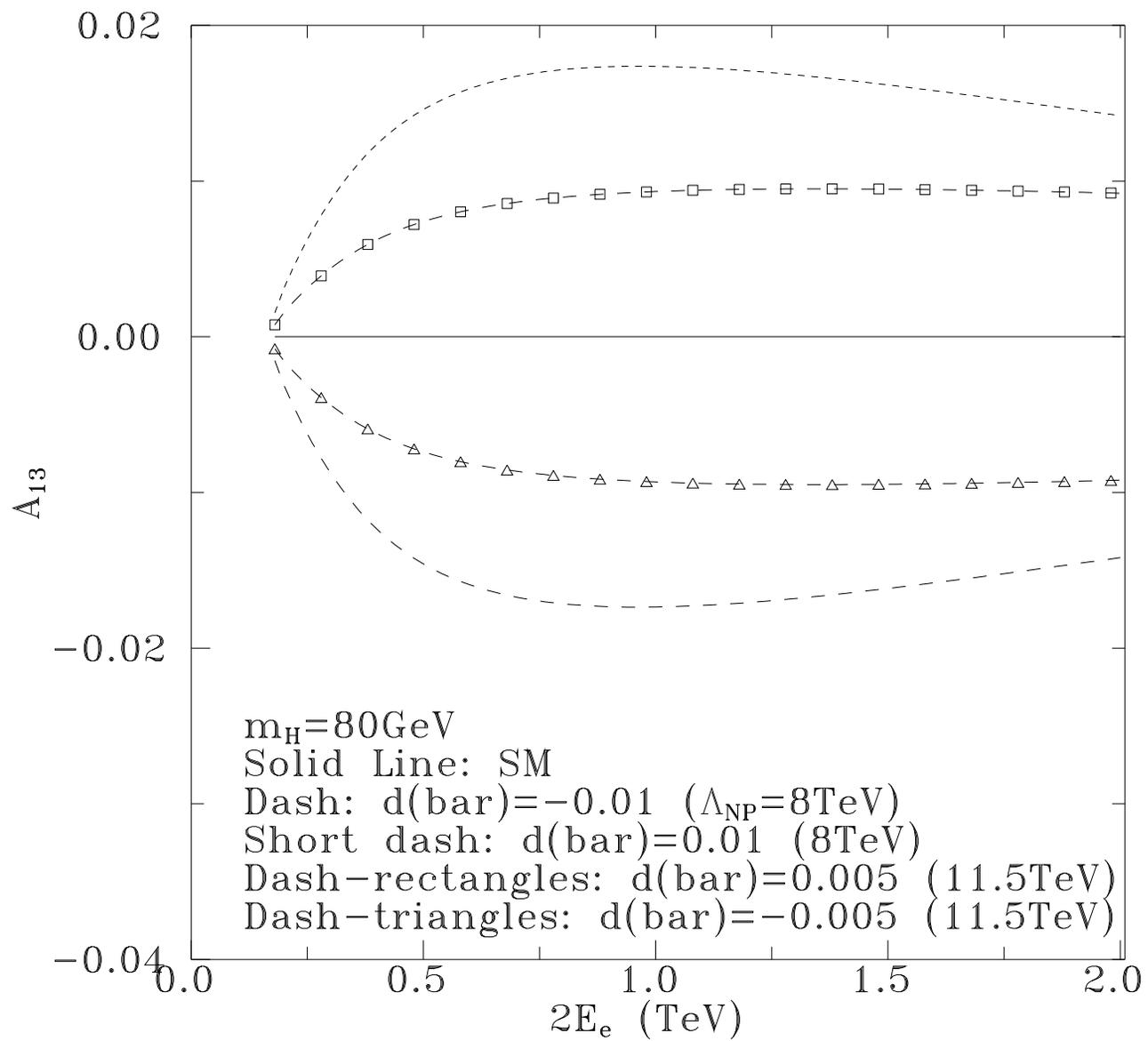
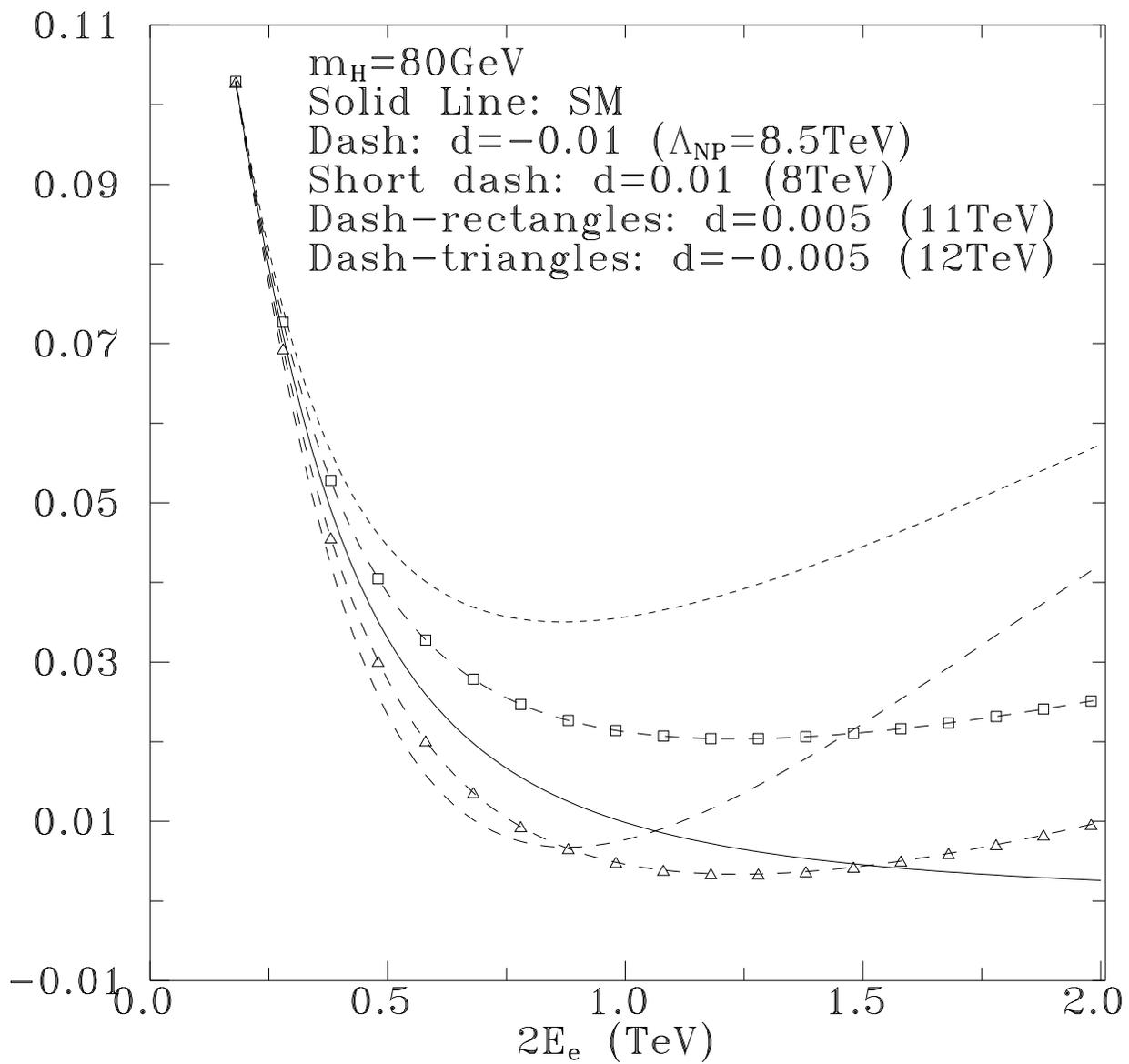
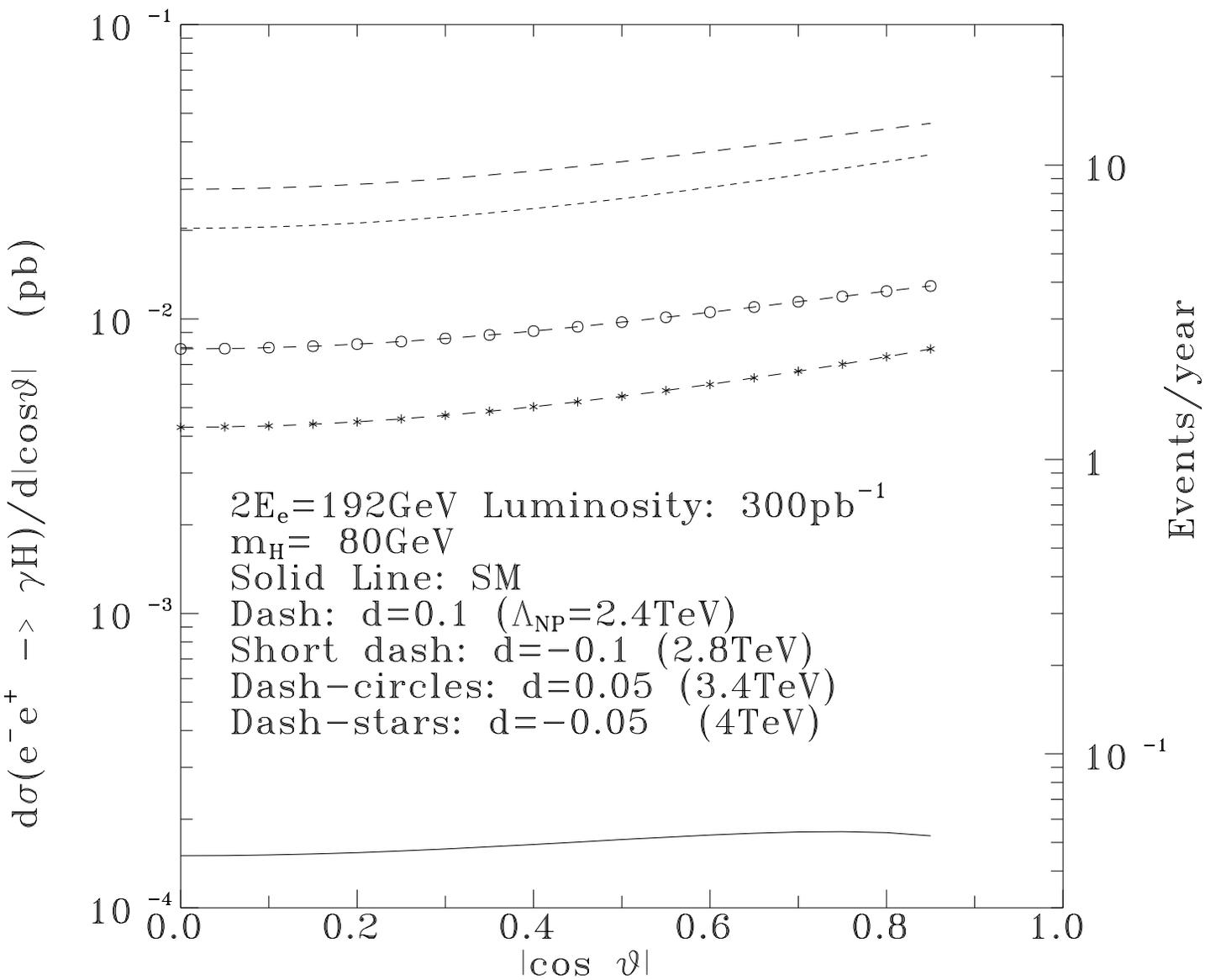
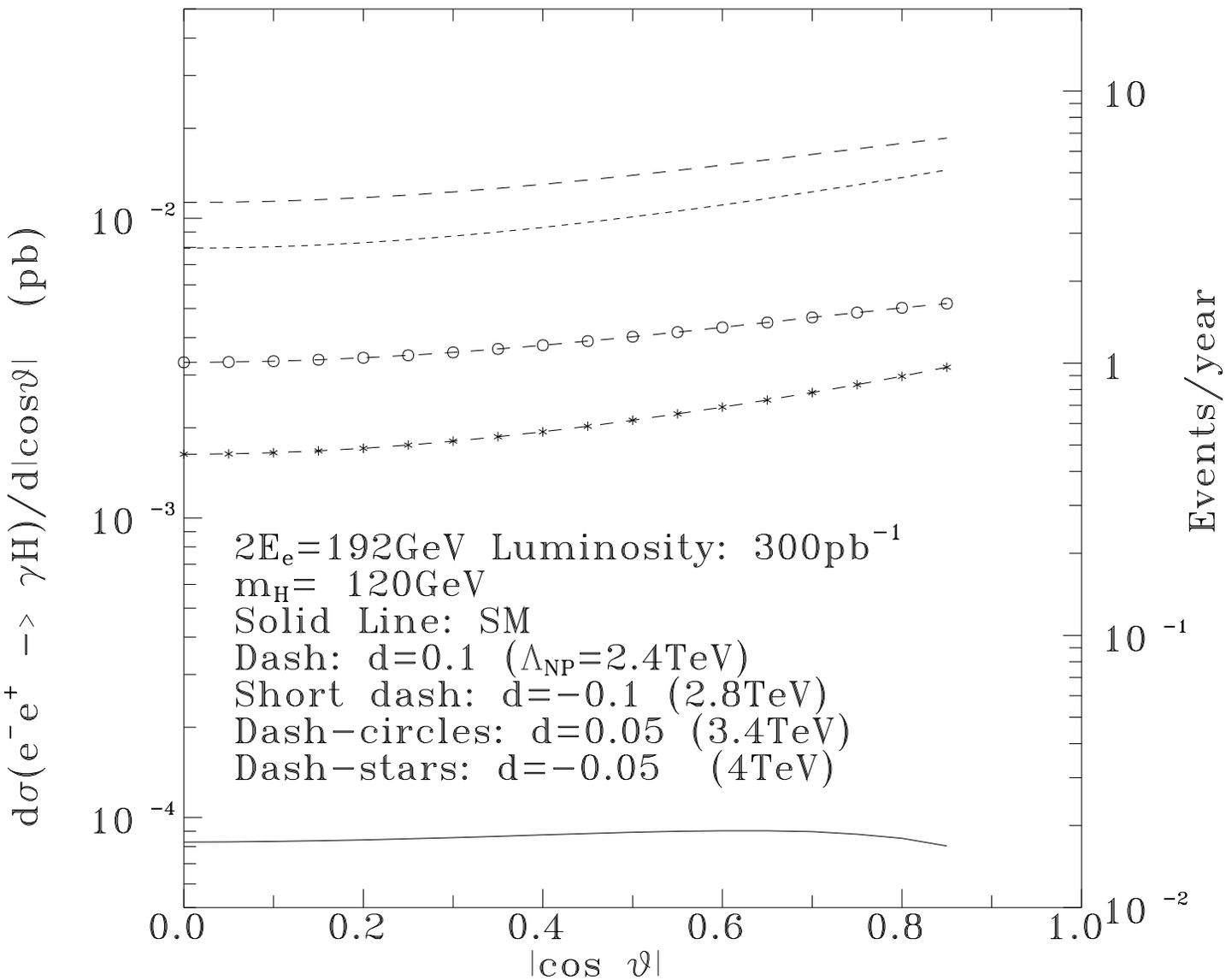
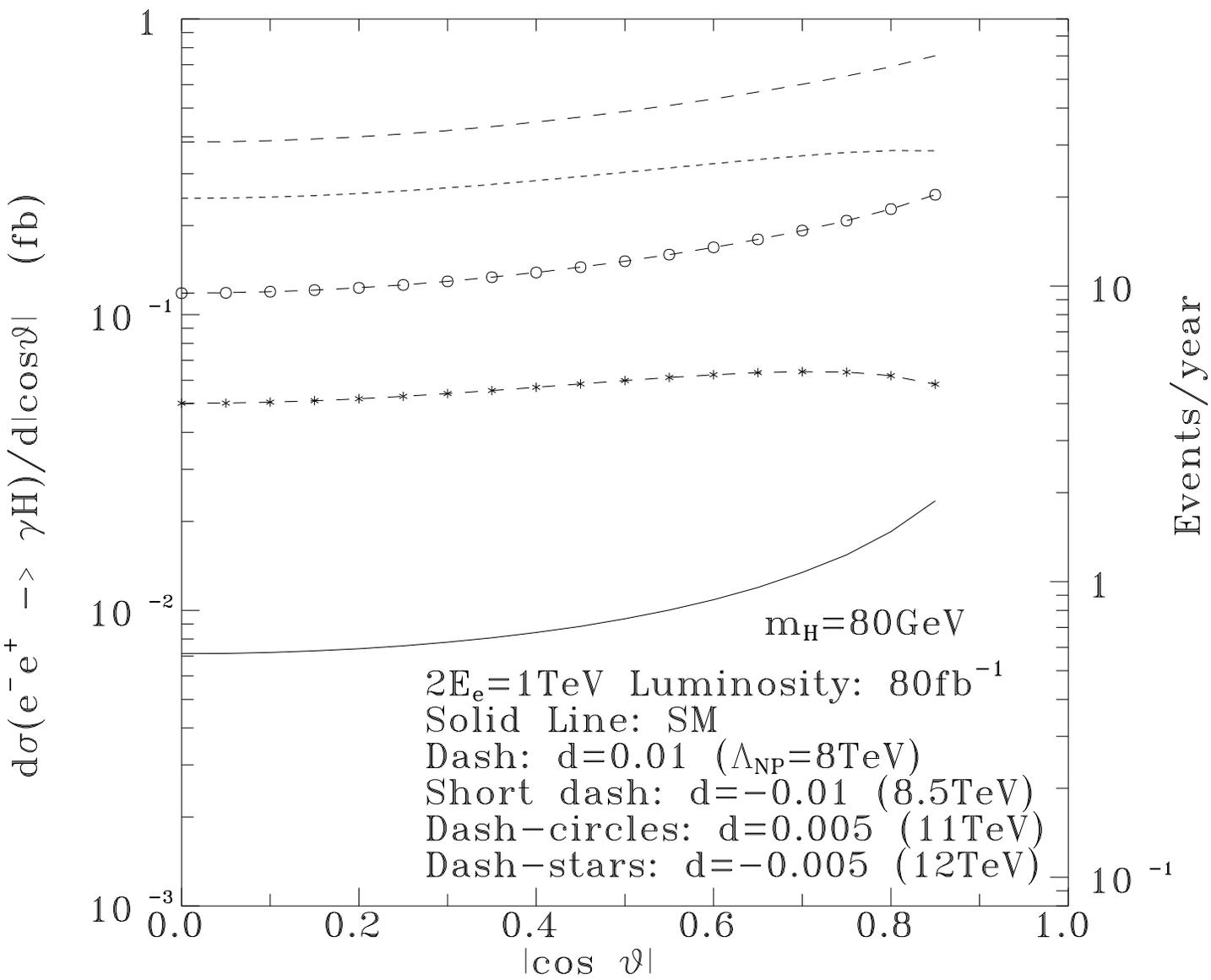
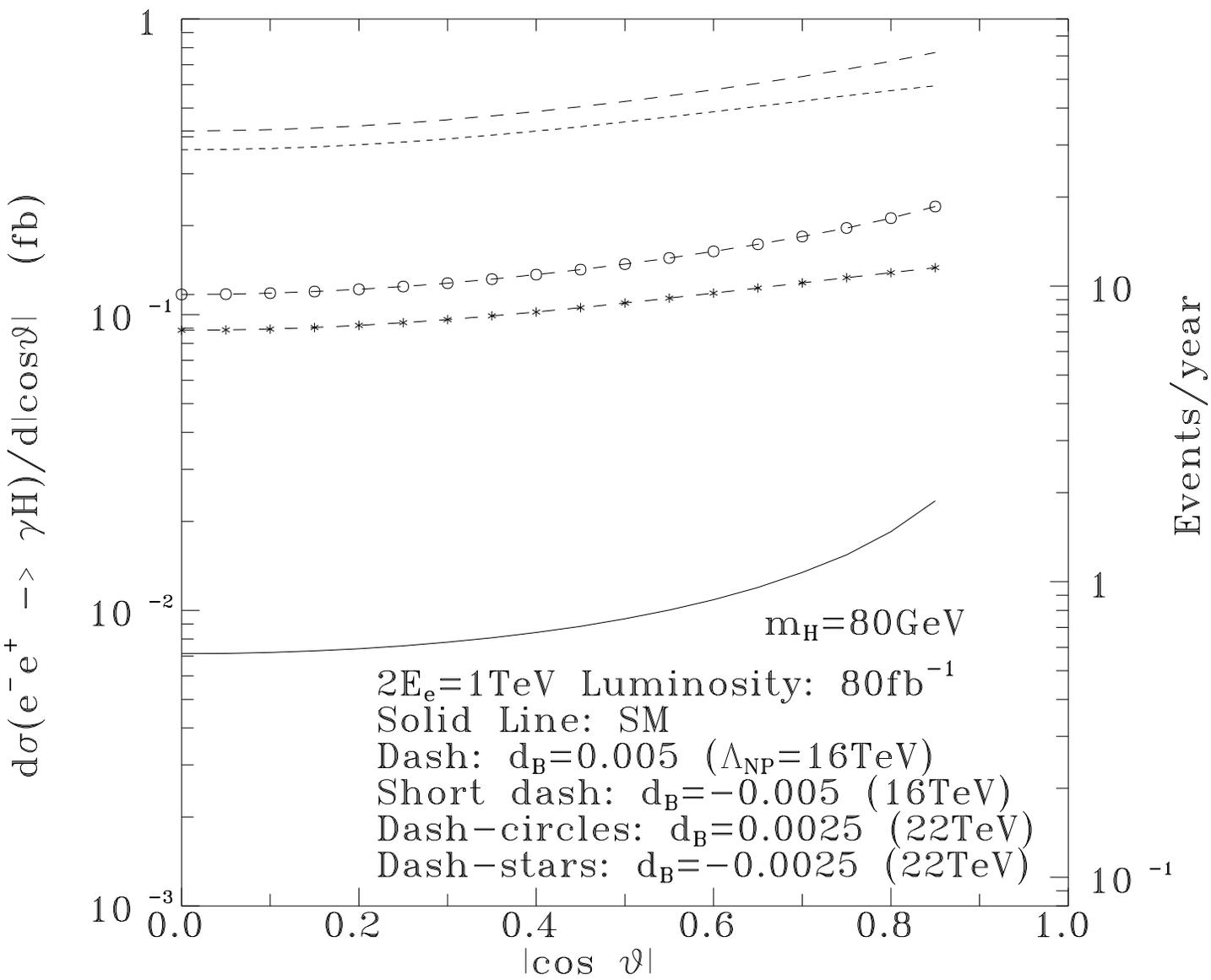
\end{document}